\documentclass[aps,prl,floatfix,amsmath,amssymb,showpacs,unsortedaddress,superscriptaddress,twocolumn]{revtex4-1}

\usepackage{graphicx}
\usepackage{amsmath}
\usepackage{amssymb}
\usepackage{amscd}
\usepackage{bm}
\usepackage{enumerate}
\usepackage{type1cm}
\usepackage{lettrine}
\usepackage{mathrsfs}
\usepackage{calrsfs}
\usepackage{epsfig}
\usepackage{subfigure}
\usepackage{psfrag}

\def\beq{\begin{equation}}
\def\eeq{\end{equation}}
\def\bea{\begin{eqnarray}}
\def\eea{\end{eqnarray}}

\newcommand{\mean}[1]{\langle#1\rangle}

\newcommand{\phase}{\phi}

\renewcommand{\l}{\ell}
\newcommand{\lst}{\l^{st}}
\newcommand{\lstmin}{\l^{st}_\text{min}}
\newcommand{\lsw}{\l^{sw}}
\newcommand{\lswmin}{\l^{sw}_\text{min}}

\newcommand{\Rst}{R^{st}}
\newcommand{\Rsw}{R^{sw}}

\newcommand{\wmean}{\bar{w}}
\newcommand{\wmin}{w_\text{min}}
\newcommand{\wmax}{w_\text{max}}
\newcommand{\wstmax}{w^{st}_\text{max}}
\newcommand{\wstmin}{w^{st}_\text{min}}
\newcommand{\wswmax}{w^{sw}_\text{max}}
\newcommand{\wswmin}{w^{sw}_\text{min}}

\makeatletter
\renewcommand*{\@fnsymbol}[1]{\ensuremath{\ifcase#1\or \dagger\or *\or \ddagger\or
   \mathsection\or \mathparagraph\or \|\or **\or \dagger\dagger
   \or \ddagger\ddagger \else\@ctrerr\fi}}
\makeatother

\begin{document}

\title{Intergenerational continuity of cell shape dynamics in \textit{Caulobacter crescentus}}
\author{Charles S. Wright}
\thanks{These authors contributed equally to this work.}
\affiliation{James Franck Institute, The University of Chicago, Chicago IL 60637}
\affiliation{Institute for Biophysical Dynamics, The University of Chicago, Chicago IL 60637}
\author{Shiladitya Banerjee}
\thanks{These authors contributed equally to this work.}
\affiliation{James Franck Institute, The University of Chicago, Chicago IL 60637}
\author{Srividya Iyer-Biswas}
\affiliation{James Franck Institute, The University of Chicago, Chicago IL 60637}
\affiliation{Institute for Biophysical Dynamics, The University of Chicago, Chicago IL 60637}
\author{Sean Crosson}
\affiliation{Institute for Biophysical Dynamics, The University of Chicago, Chicago IL 60637}
\affiliation{Department of Biochemistry and Molecular Biology, The University of Chicago, Chicago IL 60637}
\author{Aaron R. Dinner}
\altaffiliation{To whom correspondence may be addressed. Email: dinner@uchicago.edu or nfschere@uchicago.edu}
\affiliation{James Franck Institute, The University of Chicago, Chicago IL 60637}
\affiliation{Institute for Biophysical Dynamics, The University of Chicago, Chicago IL 60637}
\affiliation{Department of Chemistry, The University of Chicago, Chicago IL 60637}
\author{Norbert F. Scherer}
\altaffiliation{To whom correspondence may be addressed. Email: dinner@uchicago.edu or nfschere@uchicago.edu}
\affiliation{James Franck Institute, The University of Chicago, Chicago IL 60637}
\affiliation{Institute for Biophysical Dynamics, The University of Chicago, Chicago IL 60637}
\affiliation{Department of Chemistry, The University of Chicago, Chicago IL 60637}

\begin{abstract}
We investigate the intergenerational shape dynamics of single \textit{Caulobacter crescentus} cells using a novel combination of imaging techniques and theoretical modeling. We determine the dynamics of cell pole-to-pole lengths, cross-sectional widths, and medial curvatures from high accuracy measurements of cell contours. Moreover, these shape parameters are determined for over 250 cells across approximately 10000 total generations, which affords high statistical precision. Our data and model show that constriction is initiated early in the cell cycle and that its dynamics are controlled by the time scale of exponential longitudinal growth. Based on our extensive and detailed growth and contour data, we develop a minimal mechanical model that quantitatively accounts for the cell shape dynamics and suggests that the asymmetric location of the division plane reflects the distinct mechanical properties of the stalked and swarmer poles.  Furthermore, we find that the asymmetry in the division plane location is inherited from the previous generation. We interpret these results in terms of the current molecular understanding of shape, growth, and division of \textit{C.\ crescentus}.
\end{abstract}

\maketitle

Cell shape both reflects \cite{cabeen2005} and regulates \cite{huang2000} biological function.  The importance of cell shape is exemplified by bacteria, which rely on specific localization of structural proteins for spatiotemporal organization \cite{laloux2014}.  Bacteria take forms resembling spheres, spirals, rods, and crescents.  These shapes are defined by cell walls \cite{koch2001} consisting of networks of glycan strands cross-linked by peptide chains to form a thin peptidoglycan meshwork \cite{gan2008}.  Super-resolution imaging is now revealing the internal positions of associated proteins \cite{gahlmann2014}.  These include cytoskeletal proteins such as MreB, a homolog of actin \cite{wachi1987,iwai2002,jiang2011,figge2004},  intermediate filament-like bundles of CreS (crescentin) \cite{ausmees2003,cabeen2009}, and FtsZ, a homolog of tubulin \cite{erickson2010}.  However, due to the inherently stochastic nature of molecular processes, understanding how these proteins act collectively to exert mechanical stresses and modulate the effects of turgor pressure and other environmental factors requires complementary methods such as high-throughput, quantitative optical imaging.

Multigenerational imaging data for bacterial cells can now be obtained from microfluidic devices of various designs \cite{wang2010,moffitt2012,shaw2012,long2013,campos2014}. Still, a common limitation of most devices is that the environmental conditions change throughout the course of the experiment, particularly as geometric growth of the population results in crowding of the experimental imaging spaces. We previously addressed this issue by engineering a \textit{C.\ crescentus} strain in which cell adhesion is switched on and off by a small molecule (and inducible promoter) \cite{iyer-biswas2014b}, allowing measurements to be made in a simple microfluidic device \cite{iyer-biswas2014b,siegal-gaskins2008,lin2010,lin2012}. This technology allows imaging $>$100 generations of growth of an identical set of 250--500 single cells distributed over $\sim$25 fields of view. Thus cell density is low and remains constant. These studies afforded sufficient statistical precision to show that single \textit{C.\ crescentus} cells grow exponentially in size and divide upon reaching a critical multiple ($\approx$1.8) of their initial sizes \cite{iyer-biswas2014b}. Satisfaction of a series of scaling laws predicted by a simple stochastic model for exponential growth indicates that these dynamics can be characterized by a single time scale \cite{iyer-biswas2014a,iyer-biswas2014b}.

In this paper, we use more advanced image analysis methods to extract cell shape contours from these data. The resulting geometric parameters, together with mathematical models, provide insights into growth and division in \textit{C.\ crescentus} and the plausible role of cell wall mechanics and dynamics in these processes. Specifically, we identify natural variables for tracking cell dynamics, and develop a minimal mechanical model that shows how longitudinal growth can arise from an isotropic pressure.  We then examine the dynamics of cell constriction and unexpectedly find that it is governed by the same time constant as exponential growth. This important finding can be understood in terms of an intuitive geometric model that relates the constriction dynamics to the kinetics of the growth of septal cell wall.  We further suggest that the site of constriction can arise from differences in materials properties of the poles and show that it is established in the previous generation---i.e., the location of the site of division can be predicted before formation of the divisome. We relate our results to the known dynamics of contributing molecular factors and existing models for bacterial growth and division.

\begin{figure*}[!htp]
\includegraphics[scale=0.58]{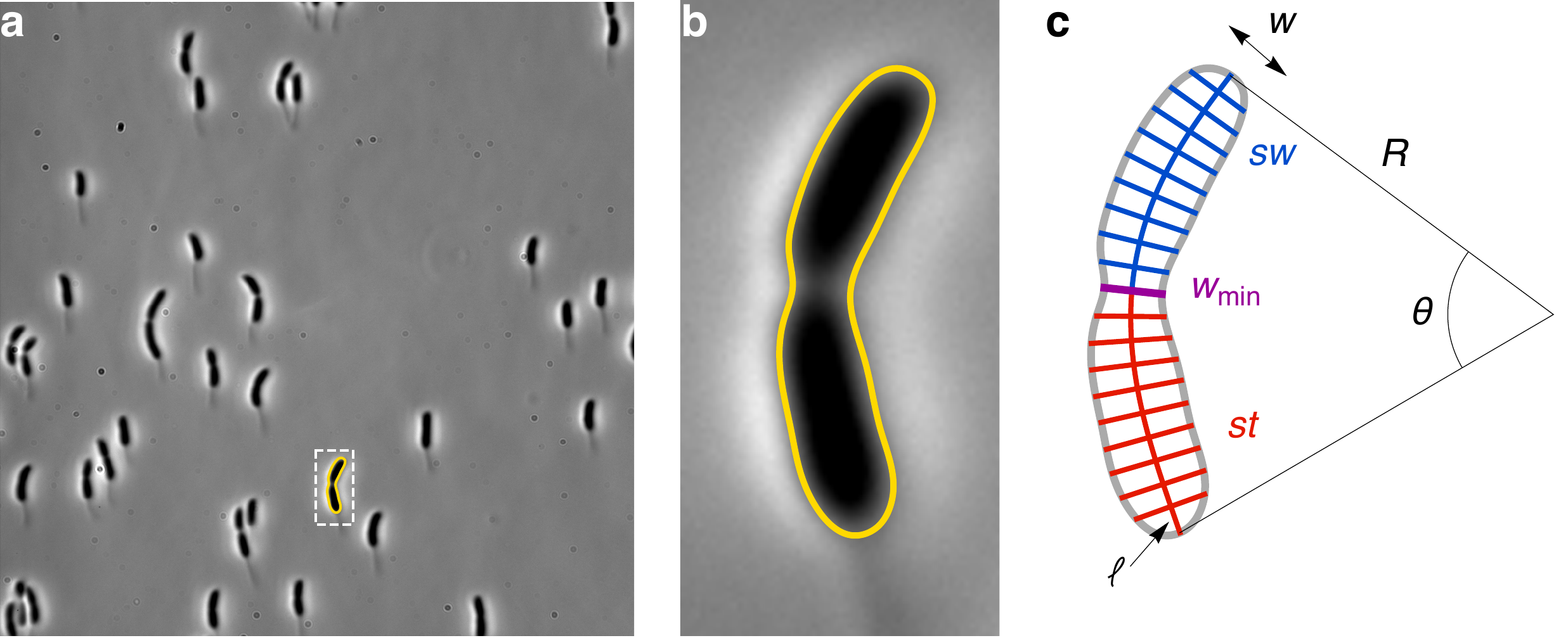}
\caption{\textbf{Determination of cell contour and definition of shape parameters.}
\textbf{(a)} A representative phase contrast image of one field of view. The solution flow in the microfluidic channel is from bottom to top. \textbf{(b)} Zoomed image of the yellow highlighted cell from \textbf{a} and its splined contour. \textbf{(c)} Schematic of a contour illustrating the shape parameters. The cell medial axis is calculated from pole to pole; it defines both cell length $\l(\phase)$ and radius of curvature $R(\phase)$, which lead directly to the spanning angle $\theta(\phase)$. The cell width $w(\phase,u)$ is a parametric quantity, calculated as the length of the rib perpendicular to the medial axis at a specified distance from the stalked pole, $u(\phase)$. The location of the global minimum of the width $\wmin(\phase)$ (purple line) can be used to segment the cell into stalked ($st$, red) and swarmer ($sw$, blue) portions.}
\label{fig:data}
\end{figure*}

\section*{Results}

\textbf{The length is sufficient to characterize the exponential growth of each cell}. 
Various techniques have been put forth to analyze cell morphology gathered from single cell images \cite{pincus2007}. Recent work on image analysis of single cells has attempted to optimize two problems: separation of distinct (but potentially overlapping) cells and accurate determination of the edge of each cell \cite{sliusarenko2011}. Because crowding is not an issue in our setup, we could focus solely on constructing an algorithm to delineate each cell contour accurately and precisely. As shown in Fig.\ 1a,b and described in the Methods section, we first segment each cell using pixel-based edge detection similar to \cite{guberman2008}, then perform spline interpolation to determine the cell contour at sub-pixel resolution. The sequence of such images for each single cell constitutes a trajectory in time $t$ that serves as the basis for quantitative analysis. Division events are then detected in an automated fashion using custom Python code, and used to divide time trajectories for each cell into individual generations. 

All data shown here were obtained by observing 260 single \textit{C.\ crescentus} cells perfused in complex medium (peptone-yeast extract; PYE) at 31$^{\circ}$C over the course of 2 days (corresponding to 9672 separate generations). Under these conditions, the mean population growth rate and division time remain constant, so we treat the trajectories of individual generations as members of a single ensemble. In other words, we segment each cell trajectory by generation and take the resulting initial frame (i.e., immediately following division) as $t=0$ minutes. In order to average over the ensemble, we then bin quantitative information according to time since division, $t$, normalized by the respective division time $\tau$. The normalized time, $\phase \equiv t / \tau$, serves as a cell-cycle phase variable.

For our quantitative analysis, we focus on a set of three intuitive and independent parameters that characterize cell shape at each stage of growth: length $\l$, width $w$, and radius of curvature $R$ (Fig.\ 1c). They are calculated directly from each splined contour as follows (see also Supplementary Fig.\ 1):
\begin{itemize}
\item{We define the length, $\l(\phase)$, as the pole-to-pole distance along the contour of the cell medial axis at the normalized time $\phase$ (Fig.\ 2a).}
\item{We assign a single radius of curvature, $R(\phase)$, to each cell based upon the best-fit circle to the medial axis (Fig.\ 2b).  Although stalked ($\Rst(\phase)$) and swarmer ($\Rsw(\phase)$) portions may be described by different radii of curvature toward the end of the cell cycle, the average radius obtained by averaging the contributions of each portion yields the same value, i.e., $\mean{(\Rst(\phase)+\Rsw(\phase))/2} \simeq \mean{R(\phase)}$ (see Supplementary Fig.\ 2c).}
\item{We define the width, $w(\phase,u)$ as the length of the perpendicular segment spanning from one side of the cell contour to the other at each position $u(\phase)$ along the medial axis, which runs from $u=0$ at the stalked pole to $u=\l$ at the swarmer pole. Furthermore, we spatially averaged the width over positions along the medial axis, $\wmean(\phase)$, to obtain a characteristic width at each time point (Fig.\ 2c).} 
\end{itemize}

The mean division time is $\mean{\tau} =73\pm7$ min, where $\mean{...}$ indicates a population average. We find that $\mean{\l(\phase)}$ increases exponentially with time constant $\mean{\kappa}^{-1}=125\pm8$ min, essentially the same time constant that we previously observed for the cross-sectional area \cite{iyer-biswas2014b}, while $\mean{\wmean(\phase)}$ and $\mean{R(\phase)}$ remain approximately constant for $0<\phase<0.5$ and each shows a dip for $0.5<\phase<0.9$ when cell constriction becomes prominent. The sharp rise in $\mean{R(\phase)}$ seen for $\phase>0.9$ results from independent alignment of the stalked and swarmer portions with the microfluidic flow as they become able to move independently (i.e., fluctuate easily about the plane of constriction). These observations confirm the assumptions in \cite{iyer-biswas2014b} that the length is sufficient to describe the growth of cell size.  Moreover, we can track the dynamics of the spanning angle, $\theta$, using the relation $\l = R\theta$.

\begin{figure*}[!htp]
\includegraphics[scale=0.58]{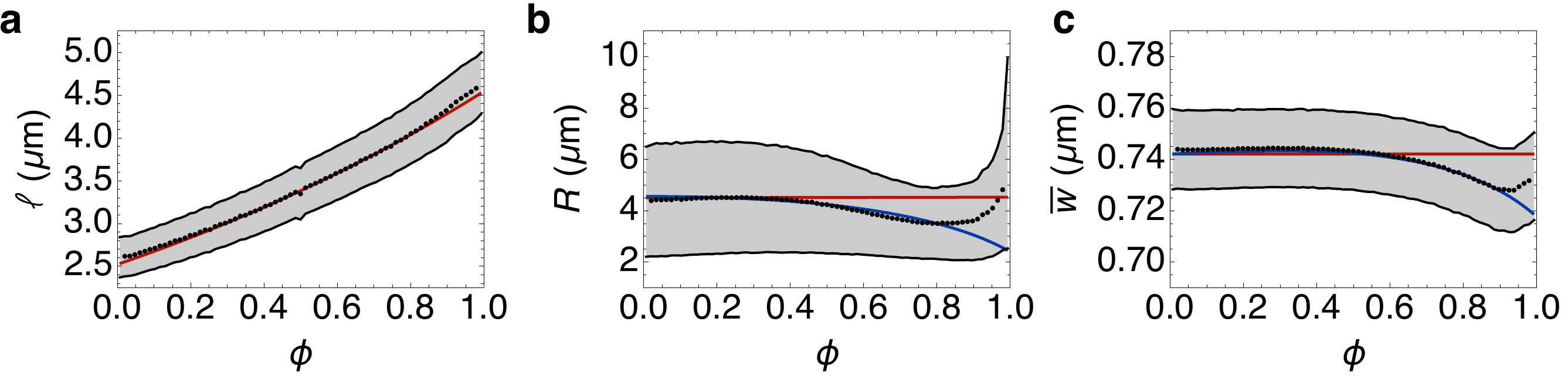}
\caption{\textbf{Dynamics of cell shape parameters.}
\textbf{(a)} Length of the cell medial axis (data shown in black and exponential fit from our mechanical model in red). \textbf{(b)} Radius of curvature of the cell medial axis, obtained by calculating the best-fit circle to the entire cell (black points), with a time-averaged mean $\mean{R}=4.44 \pm 2.12\ \mu$m ($0<\phase<0.5$). The mean-field model predicts a constant steady-state value for $\mean{R(\phase)}$ (red solid line), whereas by accounting for constriction dynamics, the model captures the dip in $\mean{R(\phase)}$ seen for $0.5<\phase<0.9$ (blue solid line). \textbf{(c)} Characteristic cell width, obtained by spatially averaging the width at each time point (black points), with a time-averaged mean $\mean{\wmean}=0.74 \pm 0.02\ \mu$m ($0<\phase<0.5$). The mean field model predicts a constant steady-state value for $\mean{\wmean(\phase)}$ (red solid line), whereas cell constriction accounts for the dip in $\mean{\wmean(\phase)}$ seen for $\phase > 0.5$ (blue solid line). The shaded regions represent $\pm 1$ standard deviation. Model parameters: $P=0.3$ MPa, $\gamma=50$ nN/$\mu$m, $k_c=2$ nN$\mu$m$^2$, $R_c=0.5$ $\mu$m, $k_m=40$ nN$\mu$m, $R_m=0.31$ $\mu$m, $\tau=73$ min.}
\label{fig:analysis}
\end{figure*}

\textbf{Mechanical model for cell shape and growth}. There are many details of cell growth and shape that require interpretation. For example, it is not obvious {\it a priori} that growth should be almost exclusively longitudinal. Therefore, we have developed a minimal mechanical model that can explain these observations. We parametrize the geometry of the cell wall by a collection of shape variables $\{q_i(t)\}$, where $q_1=\mean{R}$, $q_2=\mean{\wmean}$, and $q_3=\mean{\theta}$ are the parameters introduced above (Fig.\ 1c). As the cell grows in overall size, we postulate that the rate of growth in the shape parameter $q_i(t)$ is proportional to the net decrease in cell wall energy, $E(\{q_i(t)\})$, per unit change in $q_i(t)$~\cite{lan2007,jiang2010}. Assuming linear response, the configurational rate of strain, $\frac{1}{q_i(t)}\frac{dq_i}{dt}$, is proportional to the corresponding driving force $F_i=-\partial E/\partial q_i$, in analogy with the constitutive law of Newtonian flow~\cite{batchelor2000}:

\begin{equation}\label{eq:growth}
\frac{1}{q_i}\frac{d q_i}{dt}=\Phi_i F_i\;,
\end{equation}
where the constant $\Phi_i$ describes the rate of irreversible flow corresponding to the variable $q_i(t)$. According to equation~\eqref{eq:growth}, exponential growth occurs if $F_i$ is constant, whereas $q_i(t)$ reaches a steady-state value if $F_i(q_i)=0$ along with the condition $\partial F_i/\partial q_i<0$.  It thus remains to specify the form of $E$.

For a \textit{C.\ crescentus} cell of total volume $V$ and surface area $A$, our model for the total energy in the cell wall is given by
\begin{equation}\label{eq:energy}
E(R,\wmean,\theta) = -PV + \int dA\ \gamma + E_\text{width}+E_\text{cres}+ E_\text{div}\;,
\end{equation}
where $P$ is a constant pressure driving cell wall expansion; $\gamma$ is the tension on the surface of the cell wall; $E_\text{width}$ is the energy required to maintain the cell width; $E_\text{cres}$ represents the mechanical energy required to maintain the crescent cell shape; $E_\text{div}$ is the energy driving cell wall constriction. Traditionally $P$ was taken to be the turgor pressure~\cite{lan2007}; while the importance of the turgor pressure has recently been questioned \cite{rojas2014}, an effective pressure must still arise from the synthesis and insertion of peptidoglycan strands that constitute the cell wall. We note that a purely elastic description of cell wall mechanics would lead to a curvature-dependent surface tension~\cite{jiang2012b}. However, if growth is similar to plastic deformation, the tension is uniform~\cite{boudaoud2003}. The effective tension in our model depends on the local surface curvatures through the energy terms $E_\text{width}$ and $E_\text{cres}$, that describe harmonic wells around preferred values of surface curvatures.

The mechanical energy for maintaining width is given by
\begin{equation}\label{eq:width}
E_\text{width} = \frac{k_m}{2}\int dA \left(\frac{1}{\bar{w}/2}-\frac{1}{R_m}\right)^2,
\end{equation}
where the constant $R_m$ is the preferred radius of curvature, $k_m$ is the bending rigidity and $dA$ is a differential area element~\cite{sun2011}. Contributions to $k_m$ can come from the peptidoglycan cell wall as well as membrane-associated cytoskeletal proteins like MreB, MreC, RodZ, etc., which are known to control cell width~\cite{wachi1987,iwai2002,jiang2011}.

In addition to maintaining a constant average width, \textit{C.\ crescentus} cells exhibit a characteristic crescent shape, which relies on expression of the intermediate filament-like protein crescentin \cite{ausmees2003}. Although the mechanism by which crescentin acts is not known, various models have been proposed, including modulation of elongation rates across the cell wall~\cite{cabeen2009,sliusarenko2010} and bundling with a preferred curvature \cite{jiang2011b}.  We assume the latter and write the energy for maintaining the crescent shape as
\begin{equation}\label{eq:cres}
E_\text{cres}=\frac{k_c}{2}\int_0^{\l_c} du \left(c(u)-\frac{1}{R_c}\right)^2,
\end{equation}
where  $u$ is the arc-length parameter along the crescentin bundle attached to the cell wall, $c(u)$ is the local curvature, $R_c$ is the preferred radius of curvature, $\l_c$ is the contour length, and $k_c$ is the linear bending rigidity. Equation~\eqref{eq:cres} accounts for the compressive stresses generated by the crescentin bundle on one side of the cell wall, leading to a reduced rate of cell growth, according to equation~\eqref{eq:growth}. As a result, the cell wall grows differentially and maintains a non-zero curvature of the centerline. In the absence of crescentin ($k_c=0$), our model predicts an exponential decay in the cell curvature that leads to a straight morphology, consistent with previous observations~\cite{ausmees2003}.

Finally, one must also account for the energy driving cell wall constriction. Constriction proceeds via insertion of new peptidoglycan material at the constriction site. This process leads to the formation of daughter pole caps~\cite{typas2012}. We take constriction to be governed by an energy of the form $E_\text{div} = -\lambda S$, where $S$ is the surface area of the septal cell wall, and $\lambda$ is the energy per unit area released during peptidoglycan insertion.

\textbf{There exists an optimal cell geometry for a given mechanical energy}. To apply the model introduced above (equations\ \eqref{eq:growth} and \eqref{eq:energy}) to interpreting the data in Fig.\ 2, we assume a minimal cell geometry given by a toroidal segment with uniform radius of curvature $R(\phase)$, uniform cross-sectional width $\wmean(\phase)$ and the spanning angle $\theta(\phase)$. To this end, we estimate as many mechanical parameters as we can from the literature and then determine the rest by fitting our experimentally measured values. Turgor pressure in Gram-negative bacteria has been measured to be in the range $0.03 - 0.5$ MPa~\cite{reed1985,koch1987,deng2011}. We use a value for the effective internal pressure close to the higher end of the measured values for turgor pressure, $P=0.3$ MPa, in order to account for peptidoglycan insertion. We estimate the surface tension as $\gamma=50$ nN$/\mu$m (see Supplementary model section) and multiply it by the cell surface area $A(\phi)=\pi \wmean R\theta$ to obtain the cell wall surface energy. First, we neglect cell constriction (setting $E_\text{div}=0$) and assume that the crescentin structure spans the length of the cell wall (excluding the endcaps)~\cite{charbon2009}, with a contour length $\l_c(\phi)=(R-\wmean/2)\theta$. The mechanical properties of MreB and crescentin are likely similar to those of F-actin and intermediate filaments, respectively~\cite{van2001,ausmees2003}. However, due to a lack of direct measurements, we obtain the mechanical parameters $k_m$ and $k_c$ by fitting the model to the experimental data. 

As desired, we find that the total energy $E$ has a stable absolute minimum at particular values of the cross-section diameter $\wmean$ and the centerline radius $R$, given by  solution of $\partial E/\partial \wmean=\partial E/\partial R=0$ (see Supplementary Fig.\ 4).  The measured values are $\mean{\wmean}=0.74\pm0.02\ \mu$m and $\mean{R}=4.44\pm2.12\ \mu$m ($0<\phase<0.5$), and, as indicated by the red solid curves in Fig.\ 2b,c, the model reproduces them with $k_m=40$ nN$\mu$m and $k_c=2$ nN$\mu$m$^2$. While the fitted value for $k_c$ is numerically close to the estimate based on the known mechanical properties of intermediate filaments ($\sim 1.5$ nN$\mu$m$^2$), the value for $k_m$ is much higher than the bending rigidity of MreB bundles (see Supplementary Information). This indicates that $k_m$ is only determined in part by MreB and can have contributions from the cell wall.

Given stable values for $\wmean$ and $R$, growth is completely described by the dynamics of the angle variable $\theta(\phase)$. Consequently, we write the total energy in the scaling form $E(\wmean,R,\theta(\phase))=\theta(\phase) U(\wmean,R)$, with $U$ the energy density along the longitudinal direction. The condition for growth then becomes $U<0$, such that the energy is minimized for increasing values of $\theta(\phase)$. From our experimental data, the angle spanned by the cell centerline increases by an amount $\Delta\theta\simeq0.49$ during the entire cycle. Using our parameter estimates and fitting the data in Fig.\ 2, we obtain a numerical value for the energy density $U\simeq -5$ nN$\mu$m. We relate the angle dynamics to the length by
\begin{equation}
\frac{d\l}{d\phase}=\kappa\tau \l(\phase)=\frac{\kappa\tau}{\pi \wmean} A(\phase)
\end{equation}
where $\kappa=-\Phi_\theta U$ ($U<0$) is the rate of longitudinal growth, which can be interpreted as resulting from remodeling of peptidoglycan subunits with a mean current $\kappa/\pi \wmean$, across the cell surface area $A(\phase)$. From an exponential fit to the data for cell length (Fig.\ 2a), we obtain $\Phi_\theta=1.6\times10^{-3}$ (nN $\mu$m min)$^{-1}$, which gives us an estimate of the friction coefficient, $1/(\pi \bar{w}R\Phi_\theta)\simeq 61$ nN $\mu$m$^{-1}$ min, associated with longitudinal growth; e.g., MreB motion that is known to correlate strongly with the insertion of peptidoglycan strands \cite{garner2011}. Our results are consistent with previous observations of \textit{C.\ crescentus} cells with arrested division but continued growth \cite{fiebig2010}.  

\begin{figure*}[!htp]
\includegraphics[scale=0.58]{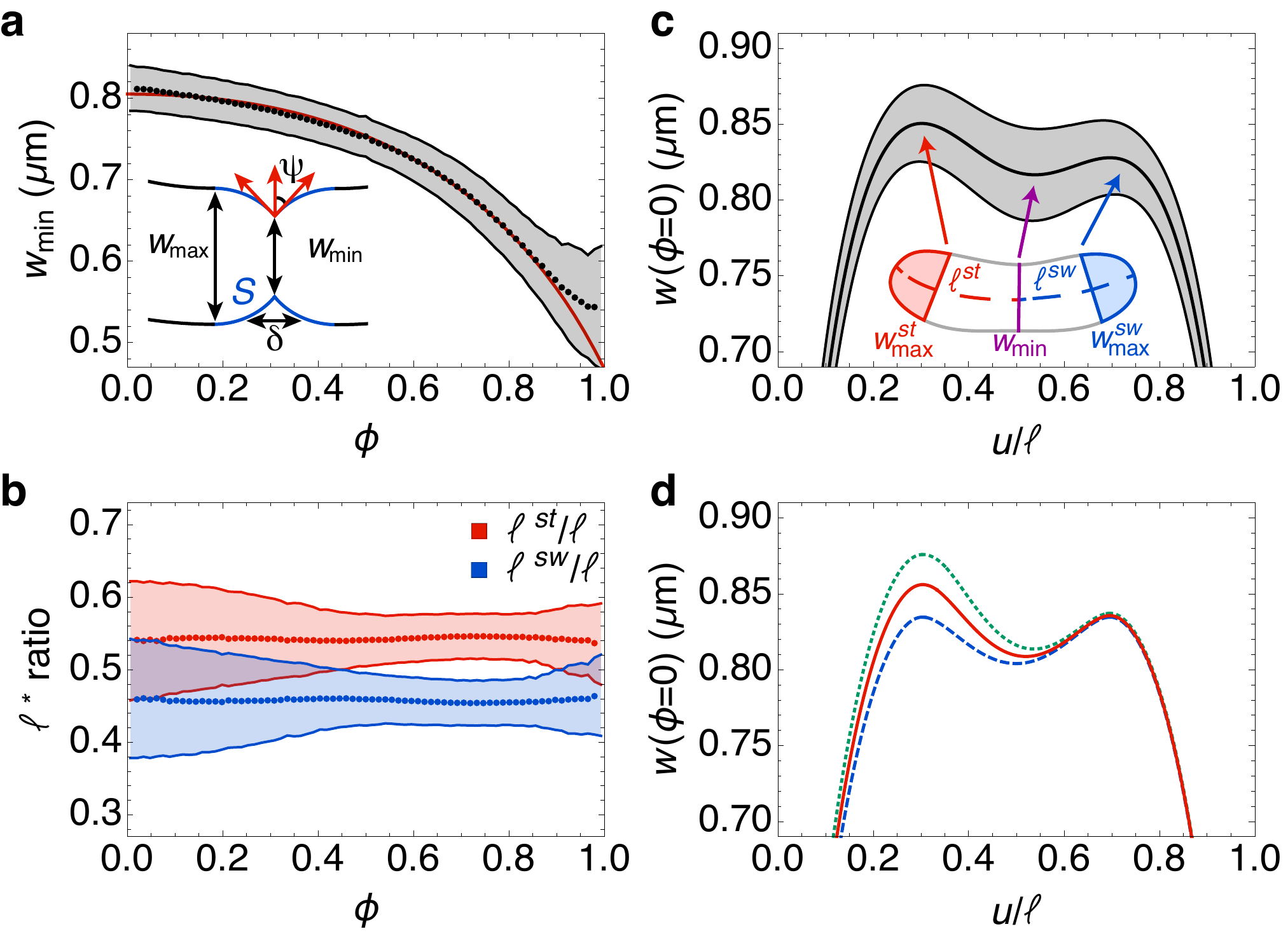}
\caption{\textbf{Timing and location of the division plane.}
\textbf{(a)} Time-dependence of $\mean{\wmin(\phase)}$, with the experimental points in black and the model fit (equations~\eqref{eq:wmin} and \eqref{eq:S}) in red. Fit values: $w_\text{max}=0.805\ \mu$m, $\kappa_d^{-1}=130.92$ min, $\kappa_0=0.016$ $\mu$m$^2$/min. Inset: Minimal geometry of a constricting cell, where $S$ (blue) is the septal cell wall synthesized during constriction, $\delta$ is longitudinal width of the constriction zone, and $\psi$ is the tangent angle at constriction. \textbf{(b)} Ratios of the length of the stalked (red) and swarmer (blue) portions divided by the total length ($\mean{\lst(\phase)/\l(\phase)}$ and $\mean{\lsw(\phase)/\l(\phase)}$, respectively), are approximately constant over the cell cycle. \textbf{(c)} Experimental width profile of \textit{C.\ crescentus} cells in the initial stage of the cycle ($\phase = 0$) after ensemble-averaging over all data. The width $\mean{w(\phase=0,u)}$ is plotted as a function of the distance from the stalked end normalized to the length of the cell, $u/\l$. Inset: A representative single cell contour immediately after division ($\phase = 0$), showing the location of the local minimum in the width ($\wmin$, purple line) as well as two local maxima ($\wstmax$, red line and $\wswmax$, blue line). These two local maxima in the width define the pole regions (shaded in red and blue, respectively). \textbf{(d)} Model width profiles of the cell showing symmetric and asymmetric location of the invagination near the midplane for different values of the ratio $\gamma_{p}^{st}/\gamma_{p}^{sw}=$1 (blue, dashed), 1.05 (red, solid) and 1.09 (green, dotted). Parameter values are the same as in Fig.\ 2 with $\gamma_{p}^{st}=5\gamma$~\cite{lan2007}. The shaded regions in \textbf{a}, \textbf{b} and \textbf{c} represent $\pm 1$ standard deviation.}
\label{fig:asymmetry}
\end{figure*}

\textbf{Constriction begins early and proceeds with the same time constant as exponential growth}. Having characterized the dynamics of growth, we now turn to constriction at the division plane.  As mentioned above, we obtain the experimental width at each point along each cell's medial axis. The typical width profile is non-uniform along its length, exhibiting a pronounced invagination near the cell center (with width $\wmin(\phase)$; Fig.\ 1c). This invagination, which ultimately becomes the division plane, is readily identifiable early in the cell cycle, even before noticeable constriction occurs. We discuss the kinetics of constriction in this section, and focus on its location later in the manuscript.  As shown in Fig.\ 3a (black points), $\mean{\wmin(\phase)}$ progressively decreases towards zero until pinching off at $\phase=1$. Due to the limited spatial resolution of our imaging (phase contrast microscopy), the pinch-off process occurring for $\phase>0.9$ could not be captured, but $\mean{\wmin(\phase)}$ at earlier times (i.e., $\phase<0.9$) is precisely determined as a function of $\phase$.

To model the dynamics of constriction, we assume as in Ref. \cite{reshes2008} (Fig.\ 3a, inset): (i) the shape of the zone of constriction is given by two intersecting and partially formed hemispheres with radii $\wmax/2$; and (ii) constriction proceeds by completing the missing parts of the hemisphere such that the newly formed cell wall surface maintains the curvature of the pre-formed spherical segments. As a result, a simple geometric formula is obtained that relates the width of the constriction zone, $\wmin(\phase)$, to the surface area $S(\phase)$ of the newly formed cell wall,
\begin{equation}\label{eq:wmin}
\wmin(\phase) = \wmax \sqrt{1-\left(S(\phase)/S_\text{max}\right)^2}\;,
\end{equation} 
where $S_\text{max} = \pi \wmax^2$ is the maximum surface area achieved by the caps as the constriction process is completed, i.e., when $\wmin(\phase=1)=0$. We assume that the addition of new cell wall near the division plane initiates with a rate, $\kappa_0$, and thereafter grows exponentially with a rate, $\kappa_d$, according to, 
\begin{equation}\label{eq:S}
\frac{1}{\tau}\frac{dS}{d\phase}=\kappa_d S(\phase) + \kappa_0 \;,
\end{equation}
subject to the initial condition $S(\phase=0)=0$. The first term on the right-hand side of equation~\eqref{eq:S} follows from equation~\eqref{eq:growth}, using $S(\phase)$ as the shape variable, after incorporating the constriction energy $E_\text{div}(\phase)$. The rate of septal peptidoglycan synthesis, $\kappa_d$, is thus directly proportional to the energy per unit area released during constriction, $\lambda$. The solution, $S(\phase) = \kappa_0(e^{\kappa_d \tau\phase}-1)/\kappa_d$, can then be substituted into equation~\eqref{eq:wmin} to derive the time-dependence of $\wmin(\phase)$, whose dynamics is controlled by two time scales: $\kappa_d^{-1}$ and $S_\text{max}\kappa_0^{-1}$.

Fitting equation~\eqref{eq:wmin} with the data for $\mean{\wmin(\phase)}$, we obtain $\kappa_d^{-1}\simeq131$ min and $S_\text{max}\kappa_0^{-1}\simeq 118$ min. The fitted values for the time constants controlling constriction dynamics ($\kappa_d^{-1}$ and $S_\text{max}\kappa_0^{-1}$) are remarkably similar to that of exponential cell elongation ($\mean{\kappa}^{-1}\simeq 125$ min). This shows that septal growth proceeds at a rate comparable to longitudinal growth. Therefore, one of the main conclusions that we draw is that cell wall constriction (Fig.\ 3a) is controlled by the same time constant as exponential longitudinal growth (Fig.\ 2a).

Having determined the dynamics of $\wmin(\phase)$, we compute the average width across the entire cell $\wmean(\phase)$ using the simplified shape of the constriction zone as shown in Fig.\ 3a (inset). The resultant prediction (blue solid curve in Fig.\ 2c) is in excellent agreement with the experimental data and captures the dip in $\mean{\wmean(\phase)}$ seen for $\phase>0.5$. Constriction also leads to a drop in the average radius of curvature of the centerline, as shown by the experimental data in Fig.\ 2b. In the supplementary material we derive a relation between the centerline radius of curvature $R(\phase)$ and the minimum width $\wmin(\phase)$, given by $R^{-1}(dR/d\phi)=\wmin^{-1}(d\wmin/d\phi)$, predicting that cell curvature increases at the same rate as $\wmin(\phase)$ drops. Using this relation, we are able to quantitatively capture the dip in $\mean{R(\phase)}$ seen for $\phase>0.5$ (solid blue curve in Fig.\ 2b) without invoking any additional fitting parameters.

\textbf{Origin of the asymmetric location of the primary invagination}. We now consider the position of the division plane and its interplay with cell shape.  As shown in Fig.\ 3b, the distance of the width minimum from the stalked pole ($\lst(\phase)$) increases through the cell cycle at the same rate as the full length of the growing cell ($\l(\phase)$), such that their ratio remains constant with time-averaged mean $\mean{\lst/\l} = 0.54\pm0.05$. The presence of the primary invagination early in the cell cycle is reiterated in Fig.\ 3c, which shows the width profile constructed by ensemble-averaging over each cell at the timepoint immediately following division. In addition to the width minimum $\wmin(\phase)$, there are two characteristic maxima near either pole, $\wstmax(\phase)$ and $\wswmax(\phase)$, respectively (Fig.\ 3c, inset). As evident in Fig.\ 3c, the stalked pole diameter $\mean{\wstmax(\phase)}$ is on average larger than its swarmer counterpart $\mean{\wswmax(\phase)}$ (also see Supplementary Fig.\ 2a).

We show that the asymmetric location of the invagination (and the asymmetric width profile) can originate from the distinct mechanical properties inherent to the pole caps in \textit{C.\ crescentus}. The shapes of the cell poles can be explained by Laplace's law that relates the pressure difference, $P$, across the cell wall to the surface tensions in the stalked or the swarmer pole, $\gamma_{p}^{st,sw}$. The radii of curvature of the poles then follow from Laplace's law
\begin{equation}\label{eq:laplace}
R_p^{st,sw}=\frac{2\gamma_p^{st,sw}}{P}\;,
\end{equation}
where the superscript ($st,sw$) denotes the stalked or the swarmer pole. Thus a larger radius of curvature in the poles has to be compensated by a higher surface tension to maintain a constant pressure difference $P$. Assuming that the poles form hemispheres, we have $R_p^{st,sw}=w_\text{max}^{st,sw}/2$. Our data indicate that the early time ratio for $\wstmax(\phase)/\wswmax(\phase)$ ($\phase<0.1$) shows a strong positive correlation with the ratio $\lst(\phase)/\lsw(\phase)$, with an average value $\mean{\wstmax/\wswmax}\simeq1.04$ (see Supplementary Fig 3a). Laplace's law then requires that the stalked pole be mechanically stiffer than the swarmer pole; $\gamma_p^{st}>\gamma_p^{sw}$. This observation suggests that the asymmetry in the lengths of the stalked and swarmer parts of the cell depends upon different mechanical properties of the respective poles.

To quantitatively support this claim, we investigate an effective contour model for the cell shape. To this end, we assume that the fluctuations in cell shape relax more rapidly than the time scale of growth. This separation of timescales allows us to derive the equation governing the cell contour by minimizing the total mechanical energy (equation\ \eqref{eq:energy}). From the solution we compute the resultant width profile for the entire cell (see Supplementary model section). As shown in Fig.\ 3d, the model with asymmetric surface tensions of the poles causes the primary invagination to occur away from the cell mid-plane. The spatial location of the invagination relative to the cell length depends linearly on the ratio $\gamma_p^{st}/\gamma_p^{sw}$. Symmetry is restored for $\gamma_p^{st}/\gamma_p^{sw}=1$, as shown in Fig.\ 3d (blue dashed curve). We note that a gradient in $\gamma$ along the cell body would imply differences in longitudinal growth rates between the stalked and the swarmer portions of the cell (Eq.~\eqref{eq:growth}). Our data exclude this possibility since both $\lst(\phase)$ and $\lsw(\phase)$ grow at the same rate $\kappa$, as evidenced by the constancy of their ratio (Fig.\ 3b and Supplementary Fig.\ 2d). Because \textit{C. crescentus} does not exhibit polar growth, the \textit{polar stiffness model} is consistent with the observed uniformity in longitudinal growth rate. In addition, the non-uniformity in cell width comes from the differences in mechanical response in the cell wall due to preferential attachment of crescentin along the concave sidewall. For a creS mutant cell (where $k_c=0$), our model predicts a uniform width profile before the onset of constriction.

\begin{figure*}[!htp]
\includegraphics[scale=0.58]{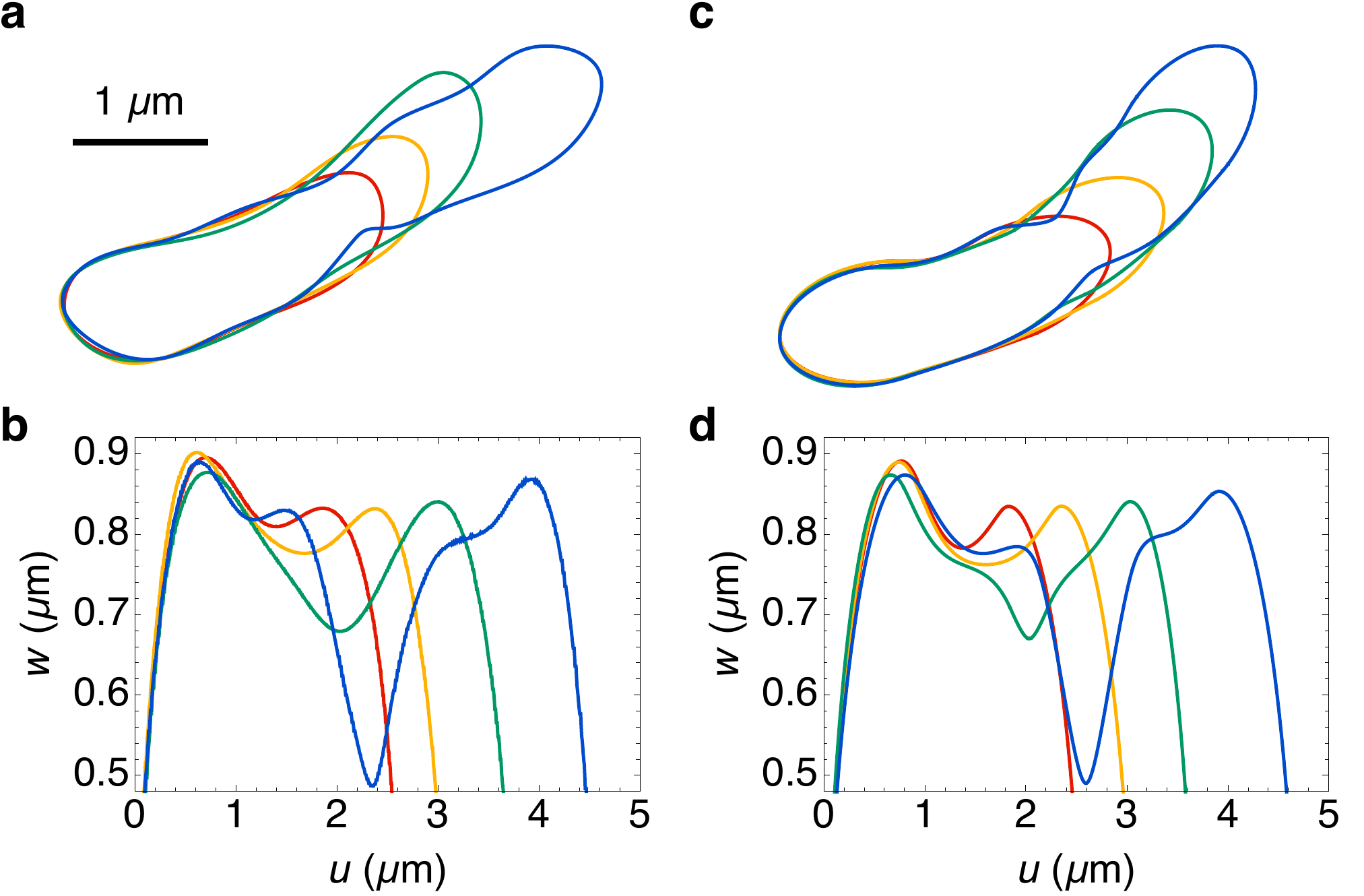}
\caption{\textbf{Comparison of experimental and model cell contours and width profiles.}
\textbf{(a)} Splined contours of a growing and constricting cell at different values of normalized time $\phase=0.0$ (red), $0.33$ (orange), $0.67$ (green) and $1.0$ (blue). \textbf{(b)} Experimental width profiles plotted against absolute distance from the stalked pole, corresponding to contours in \textbf{a}. \textbf{(c)} Contours computed from the cell shape model at different values of $\wmin$ and $\l$ corresponding to the time points in \textbf{a}. \textbf{(d)} Model width profiles corresponding to the contours in \textbf{c}.}
\label{fig:constriction}
\end{figure*}

\begin{figure*}[!htp]
\includegraphics[scale=0.58]{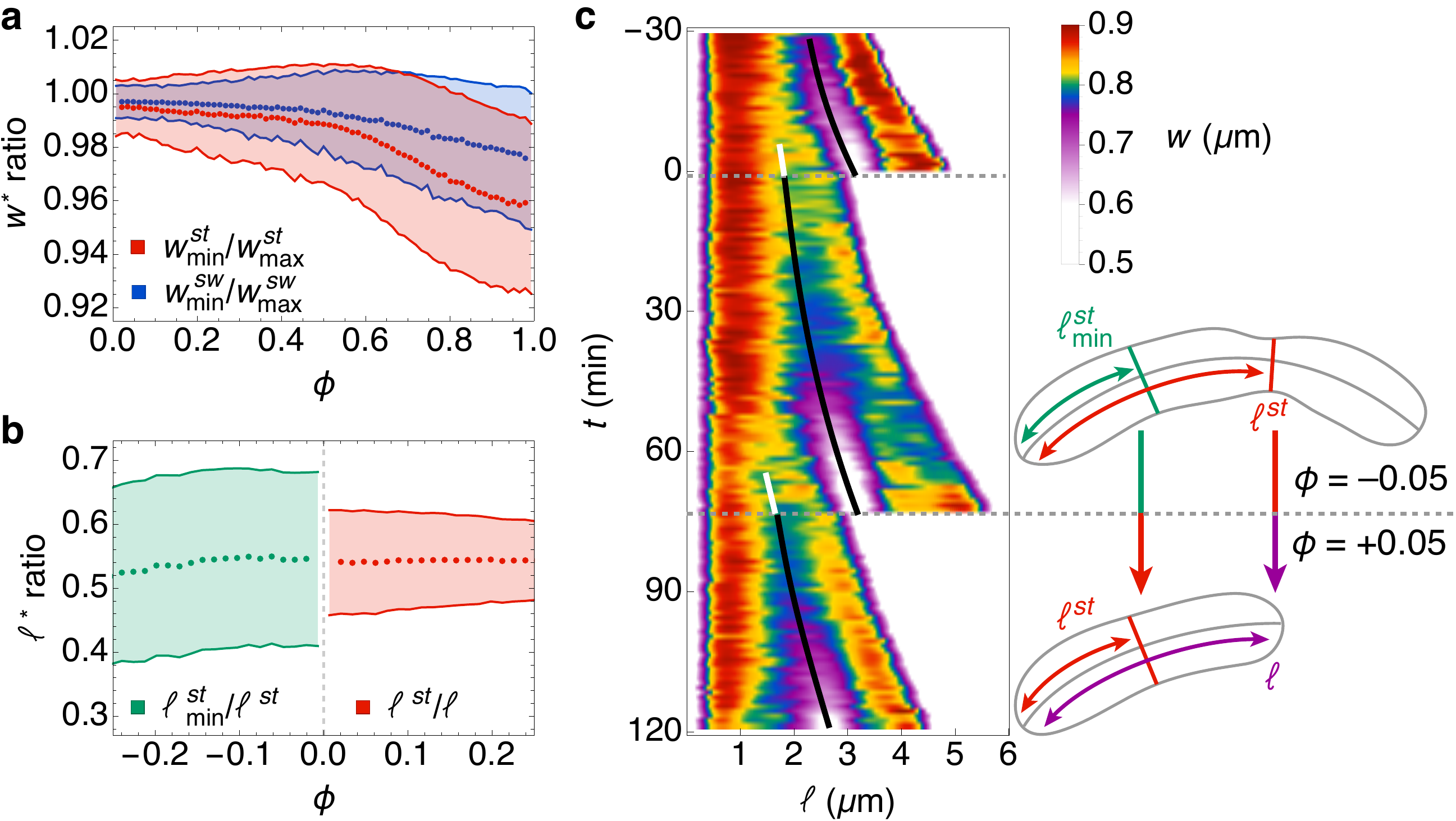}
\caption{\textbf{Location of division plane is set in the previous generation.}
\textbf{(a)} Value of the minimum width normalized by the respective maximum width for the stalked ($\mean{\wstmin(\phase)/\wstmax(\phase)}$) and the swarmer ($\mean{\wswmin(\phase)/\wswmax(\phase)}$) parts. \textbf{(b)} Ratio of length from stalked pole to secondary minimum normalized by length from stalked pole to primary minimum ($\mean{\lstmin(\phase)/\lst(\phase)}$, green) and ratio of length from stalked pole to primary minimum normalized by total length ($\mean{\lst(\phase)/\l(\phase)}$, red). The former ratio remains constant at $\simeq$ 0.55, while the latter obtains this value at the end of the cell cycle. In comparing averages between two generations here, we indicate values of $\phase$ from the first generation as negative (e.g., $\phase=-0.2$ is 20\% of the way from the subsequent division). The green points are not shown for $\phase<-0.25$ due to increased errors in identification. \textbf{(c)} Kymograph of width profiles for a typical cell over two generations. The time evolution of the widths (color scale) illustrates continuity of the location of the minima across generations. That is, the location of the secondary minimum just before division ($\lstmin$, white line) becomes the primary minimum ($\lst$, black line) just after division (horizontal dashed line). The schematic at right shows two measured contours that correspond to time points immediately before ($\phase=-0.05$) and after ($\phase=+0.05$) the division event shown in the kymograph.}
\label{fig:invagination}
\end{figure*}

\textbf{Cell shape evolution during wall constriction}. The experimental width profiles show that the growing and constricting cells typically develop a second minimum in width (Fig.\ 4a,b). These secondary invaginations are observed in both the stalked and swarmer portions of single cells in the predivisional stage ($\phase>0.6$), although they are more common in the stalked portions (Fig.\ 5a). We show here that these secondary minima become the primary minima in each of the daughter cells. To study the dynamics of the development of the secondary minimum we introduce a new quantity, $\lstmin(\phase)$, defined as the distance from the stalked pole to the secondary minimum in the stalked part (see Fig.\ 5c, inset). We find that the ratio $\mean{\lstmin(\phase)/\lst(\phase)}$ has a mean value of $\simeq$ 0.55 at later points in the cell cycle (Fig.\ 5b), equal to the constant ratio maintained by the distance from the stalked pole to the primary minimum, $\mean{\lst(\phase)/\l(\phase)}$. In fact the kymograph of width profiles (shown over 2 generations for a representative single cell) in Fig.\ 5c demonstrates that the predivisional secondary invaginations are inherited as primary invaginations after division. This mechanism provides continuity and inheritance of the invaginations across generations and is an intrinsic element of the mechanism for cell division in \textit{C. crescentus}.

To quantitatively explain the experimental width profiles during constriction, we use our mechanical model to determine the instantaneous cell shape by minimizing the total energy (equation\ \eqref{eq:energy}) at the specified time points  (see Supplementary model section). To take constriction into account, we impose the constraint that $w(\lst,\phase)=w_\text{min}(\phase)$, where $\wmin(\phi)$ is determined by equations~\eqref{eq:wmin} and \eqref{eq:S}. In addition, we assume non-uniform materials properties in the cell wall by taking the tension in the cell poles ($\gamma_\text{p}^{st,sw}$) and the septal region to be higher than the rest of the cell. As constriction proceeds and $\wmin(\phi)$ decreases, we compute the shape of the cell contours (Fig.\ 4c) and the corresponding width profiles (Fig.\ 4d). The computed width profiles faithfully reproduce the secondary invaginations, which become more pronounced as the daughter pole caps become prominent. An example of the experimental width profiles is shown in Fig.\ 4b at evenly-spaced intervals in time for a single generation, and the corresponding model width profiles are shown in Fig.\ 4d.

We note that the experimental cell contours in the predivisional stage ($\phase>0.9$) bend away from the initial midline axis and develop an alternate growth direction (Fig.\ 4a, blue contour). These bend deformations are induced by the microfluidic flow about the pinch-off plane; the cells become increasingly ``floppy'' as the constriction proceeds. 

\section*{Discussion and Conclusions}
The consistent propagation of a specific shape through the processes of growth and division relies upon an intricate interplay between the controlled spatiotemporal expression and localization of proteins, and cytoskeletal structural elements. The high statistical precision of our measurements allows us to gain new insights into cell morphology. From precise determination of cell contours over time, we observe that a typical cell width profile is non-uniform at all times with a pronounced primary invagination appearing during the earliest stages of the cell cycle. During cell constriction, the decrease in the minimum width is governed by the same time constant as exponential axial growth (Fig.\ 3a). Furthermore, the location of the primary invagination divides the cell contour into its stalked and swarmer compartments, such that the ratio of the length of the stalked part $\l^{st}(\phase)$ to the total pole-to-pole length $\l(\phase)$ remains constant during the cycle with a mean value $\mean{\lst(\phase)/\l(\phase)}\simeq 0.55$ (Fig.\ 3b). These observations and our mechanical model lead to two important conclusions: first, \textit{the dynamics of cell wall constriction and septal growth occur concomitantly}, and second, \textit{the asymmetric location of the primary invagination can be explained by the differences in mechanical properties in the stalked and swarmer poles}. A corollary of the first conclusion is that the size ratio threshold at division occurs naturally without requiring a complex timing mechanism~\cite{iyer-biswas2014b}.

In addition to the primary septal invagination, the cell contours exhibit a pronounced secondary invagination during the predivisional stages (Fig.\ 4). Remarkably, the secondary invaginations develop at a precise location relative to the total length of the stalked compartments, $\mean{\lstmin(\phase)/\lst(\phase)} \simeq 0.55$ (Fig.\ 5b). The data thus allow a third conclusion: \textit{these secondary invaginations are inherited as primary invaginations in each of the daughter cells, directing the formation of the division plane in the next generation}. Thus, through consistent and controlled nucleation of invaginations across generations, \textit{C.\ crescentus} cells maintain a constant ratio of the sizes of stalked and swarmer daughter cells.

Our experimental observations and the parameters in the cell shape model can be related to the current molecular understanding for Gram-negative bacteria, in particular \textit{C. crescentus}. Before the onset of noticeable constriction, cell shape is dictated by the mechanical properties of the peptidoglycan cell wall in addition to various shape-controlling proteins such as MreB, MreC, RodZ and CreS. Single molecule tracking studies have revealed that MreB forms short filamentous bundles anchored to the inner surface of the cell wall and moves circumferentially at a rate much faster than the rate of cell growth~\cite{kim2006,garner2011}. \textit{In vitro} experiments show that MreB filaments can induce indentation of lipid membranes, suggesting that they may have a preferred radius of curvature~\cite{salje2011}. Thus on time scales comparable to cell growth, $E_\text{width}$ is determined in part by the energy cost of adhering MreB bundles to the cell wall (see Supplementary model section). 

Bacterial cell division is driven by a large complex of proteins, commonly known as divisomes that assemble into the Z-ring structure near the longitudinal mid-plane of the cell~\cite{erickson2010}. The Z-ring contains FtsZ protofilaments that are assembled in a patchy band-like structure~\cite{holden2014}. FtsZ protofilaments are anchored to the cell membrane via FtsA and ZipA, and play a crucial role in driving cell wall constriction~\cite{erickson1996}. During constriction, the divisome proteins also control peptidoglycan synthesis and direct the formation of new cell wall via the activity of penicillin-binding proteins (PBPs)~\cite{scheffers2004,adams2009}. Thus the divisome plays a two-fold role by concomitantly guiding cell wall constriction and growth of the septal peptidoglycan layer.  According to our model the constriction of the cell wall is driven by the synthesis of septal cell wall at a rate $\kappa_d$ ($\sim \mean{\kappa}$), which can be directly related to the activity of PBPs triggered by the divisome assembly. Furthermore, in our model it is sufficient that the divisome guide the curvature of cell wall growth in the septal region (see Fig.\ 3a, inset).

While the mechanism behind the precise asymmetric location of the division plane in \textit{C. crescentus} cells is not well understood, it is likely that the ATPase MipZ helps division site placement by exhibiting an asymmetric concentration gradient during the predivisional stage~\cite{kiekebusch2012}. MipZ activity inhibits FtsZ assembly; as a result of polar localization of MipZ, Z-ring assembly is promoted near the mid-cell~\cite{thanbichler2006}. Our cell shape model suggests that the early time asymmetric location of the primary invagination, which develops into the division plane, is controlled by the differences in surface tensions maintained in the poles. The presence of this invagination at $\phase=0$, as inherited from the secondary invaginations in the previous generation, aids in Z-ring assembly at the site of the invagination. The curvature-sensing capability of the Z-ring may be enabled by the minimization of the FtsZ polymer conformational energy that is determined by the difference between cell surface curvature and FtsZ spontaneous curvature~\cite{erickson2010,arumugam2012}. 

A higher tension in the stalked pole can be induced by asymmetric localization of polar proteins, such as PopZ, early in the cell cycle. Experiments have shown that PopZ localizes to the stalked pole during the initial phase of the cell cycle and increasingly accumulates at the swarmer pole as the cell cycle proceeds~\cite{bowman2010}. Consistent with this observation, our data show that the correlation between the pole sizes (determined by the ratio of surface tension to pressure) and the stalked and swarmer compartment lengths tend to disappear later in the cycle (Supplementary Fig.\ 3), as cell constriction proceeds. A recent experimental study also demonstrates that molecular perturbation of Clp proteases can destroy the asymmetry of cell division in \textit{C. crescentus} ~\cite{williams2014}, suggesting the interplay of subcellular protease activity with the physical properties of the cell wall.

Earlier theoretical models have predicted that a small amount of pinch-off force from the Z-ring ($\sim 8$ pN) is sufficient to accomplish division by establishing a direction along which new peptidoglycan strands can be inserted~\cite{lan2007}. In contrast, our data combined with the mathematical model allows the interpretation that \textit{the early time asymmetric invagination in the cell wall can set the direction for the insertion of new peptidoglycan strands}. Constriction results from exponential growth of surface area in the septum (at the same rate as longitudinal extension). The instantaneous cell shape is determined by minimizing the energy functional at given values of the cell size parameters.

Finally, from our estimate of the cell wall energy density $U$ ($\simeq -5$ nN$\mu$m), we predict that a net amount $\Delta\theta\vert U \vert\simeq 2.5$ nN$\mu$m of mechanical energy is used by the peptidoglycan network for cell wall growth. For a \textit{C.\ crescentus} cell of surface area $12.5-25 \mu$m$^2$, layered with glycan strands of length $\sim$5 nm and cross-linked by peptide chains with maximally stretched length $\sim$4 nm~\cite{gan2008}, there are roughly $10^6$ peptidoglycan subunits. Thus on average, each peptidoglycan subunit can consume mechanical energy of $\sim$2.4$\times 10^{-6}$ nN$\mu$m, or $\sim$0.6 $k_BT$ at a temperature $T=31^{\circ}$C. Cell wall remodeling and insertion of new peptidoglycan material can likely create defects in the peptidoglycan network~\cite{amir2012}. One thus expects cellular materials properties to change over time, as a result of these molecular scale fluctuations. Although we neglect such variations in our mean field model, it nonetheless quantitatively captures the average trends in cell shape features. In future work we plan to more closely connect the energy terms of the continuum model with molecular details.

\section*{Methods}

\textbf{Acquisition of experimental data}. Data were acquired as in Ref. \cite{iyer-biswas2014b}. Briefly, the inducibly-sticky \textit{Caulobacter crescentus} strain FC1428 was introduced into a microfluidic device and cells were incubated for one hour in the presence of the vanillate inducer. The device was placed inside a homemade acrylic microscope enclosure ($39'' \times 28'' \times 27''$) equilibrated to 31$^{\circ}$C (temperature controller: CSC32J, Omega and heater fan: HGL419, Omega). At the start of the experiment, complex medium (peptone-yeast extract; PYE) was infused through the channel at a constant flow rate of 7 $\mu$L/min (PHD2000, Harvard Apparatus), which flushed out non-adherent cells. A microscope (Nikon Ti Eclipse with the ``perfect focus'' system) and robotic XY stage (Prior Scientific ProScan III) under computerized control (LabView 8.6, National Instrument) were used to acquire phase-contrast images at a magnification of 250X (EMCCD: Andor iXon+ DU888 1k $\times$ 1k pixels, objective: Nikon Plan Fluor 100X oil objective plus 2.5X expander, lamp: Nikon C-HFGI) and a frame rate of 1 frame/min for 15 unique fields of view over 48 hours. In the present study we use a dataset consisting of 260 cells, corresponding to 9672 generations (division events).

\textbf{Analysis of single cell shape}. The acquired phase-contrast images were analyzed using a novel routine we developed (written in Python). Each image was processed with a pixel-based edge detection algorithm that applied a local smoothing filter, followed by a bottom-hat operation. The boundary of each cell was identified by thresholding the filtered image. A smoothing B-spline was interpolated through the boundary pixels to construct each cell contour. Each identified cell was then tracked over time to build a full time series. We chose to include only cells that divided for more than 10 generations in the analysis. A minimal amount of filtering was applied to each growth curve to remove spurious points (e.g., resulting from cells coming together and touching, or cells twisting out of plane). The timing of every division was verified by visual inspection of the corresponding phase contrast images, so that the error in this quantity is approximately set by the image acquisition rate of 1 frame/min.

\section*{Acknowledgments}

We gratefully acknowledge funding from NSF Physics of Living Systems (NSF PHY-1305542), NSF Materials Research Science and Engineering Center (MRSEC) at the University of Chicago (NSF DMR-1420709), the W. M. Keck Foundation and the Graduate Program in Biophysical Sciences at the University of Chicago (T32 EB009412/EB/NIBIB NIH HHS/United States). N.F.S. also thanks the Office of Naval Research (ONR) for a National Security Science and Engineering Faculty Fellowship (NSSEFF).

\section*{Author contributions}

C.S.W., S.I.B., A.R.D., and N.F.S. designed the experiments; S.B., A.R.D., and N.F.S. designed the model; C.S.W. and S.I.B. performed the experiments and observed the phenomena reported; C.S.W. designed and implemented custom software to automate cell shape image analysis; C.S.W. and S.B. analyzed the data; S.C. contributed reagents and materials; C.S.W., S.B., A.R.D., and N.F.S. wrote the manuscript; all authors discussed the results and commented on the manuscript.

\section*{Competing financial interests}

The authors declare no competing financial interests.

\onecolumngrid
\pagebreak
\begin{center}
{\bf\Large Supplemental Material}
\end{center}
\setcounter{secnumdepth}{3}  
\setcounter{equation}{0}
\setcounter{figure}{0}
\renewcommand{\thetable}{\arabic{table}}
\renewcommand{\figurename}{\textbf{Supplementary Figure}}
\renewcommand{\tablename}{\textbf{Supplementary Table}}
\renewcommand{\theequation}{S.\arabic{equation}}
\newcommand\Scite[1]{[S\citealp{#1}]}
\makeatletter \renewcommand\@biblabel[1]{[S#1]} \makeatother

\section*{Experimental Cell Shape Parameters}

A number of quantities can be immediately calculated from each splined cell contour (Supplementary Fig.\ 1a), including the cross-sectional area. The cell medial axis was determined by calculating the Voronoi diagram of the cell contour~\Scite{brandt1994}, pruning the branches (Supplementary Fig.\ 1b), and extending this skeleton to the edges of the cell contour in a manner that preserves average curvature of the medial axis (Supplementary Fig.\ 1c). The intersections between the cell medial axis and contour represent the stalked and swarmer poles, respectively. The cell length was calculated by evaluating the distance along the medial axis between either pole. The cell widths were determined by creating ribs perpendicular to the medial axis along its length and determining the distances between their intersections with opposite sides of the cell contour (Supplementary Fig.\ 1d). To calculate time-averaged quantities, we normalized trajectories for each generation by the respective division time $\tau$ (thus converting each variable to a function of $\phase \equiv t / \tau$). We then split these data into 73 bins, as $\mean{\tau}=73\pm7$ min under these conditions, and ensemble-averaged each of these bins over every generation.

\begin{figure*}[!htp]
\includegraphics[scale=0.42]{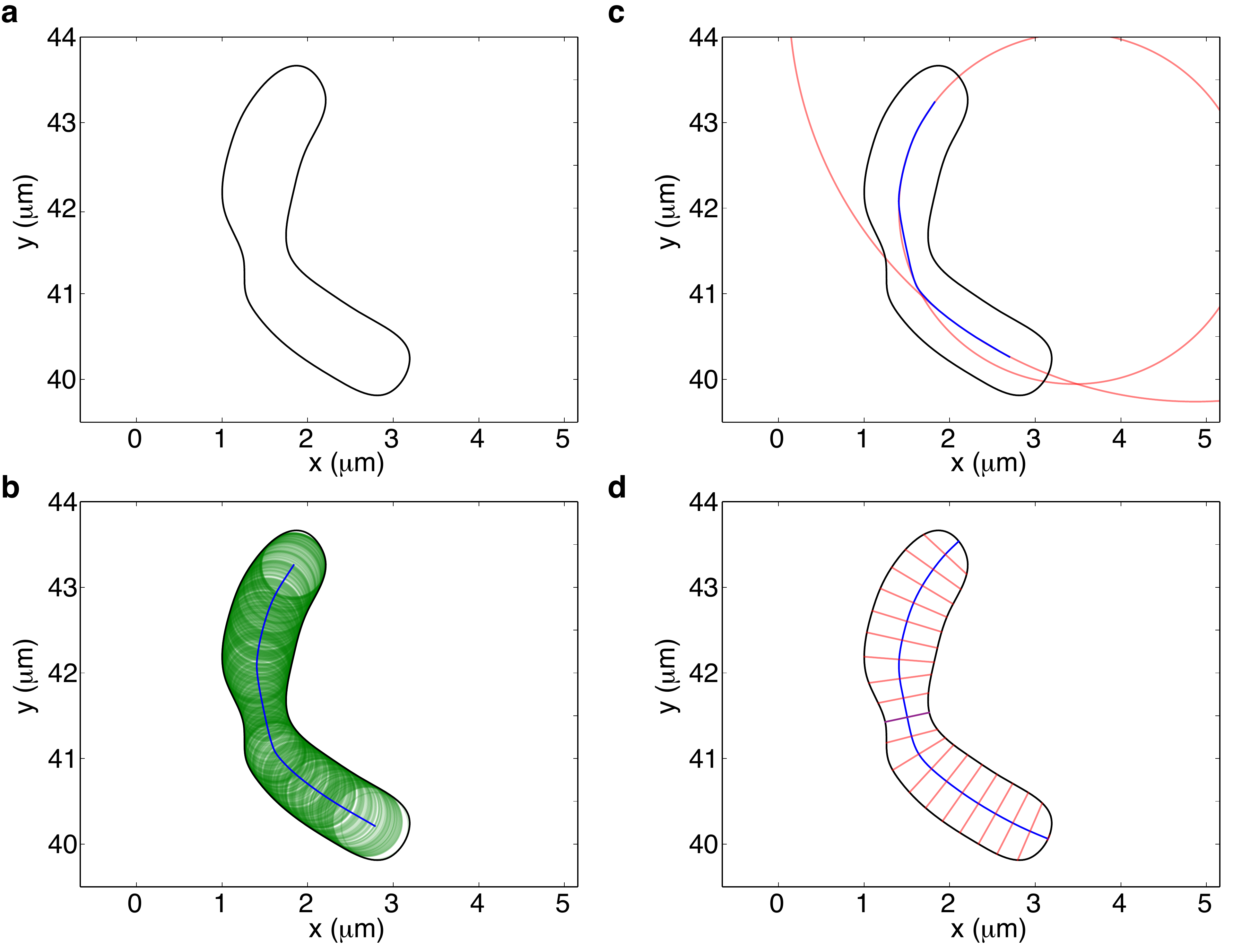}
\caption{\textbf{Procedure for calculating shape parameters}. \textbf{(a)} B-spline cell contour extracted from segmented phase contrast image. \textbf{(b)} The interior part of the medial cell axis was obtained by taking the Voronoi diagram of the cell contour, which is equivalent to the locus of inscribed circles. \textbf{(c)} The medial axis located on each side of the width minimum (excluding the constriction zone) was extended to either pole such that average radius of curvature remained constant. \textbf{(d)} The cell widths are the ribs perpendicular to the medial axis.}
\label{fig:contour}
\end{figure*}

We can define five specific values of the width according to the local minima and maxima in the width profile ($\wmin$, $\wstmax$, $\wswmax$, $\wstmin$, $\wswmin$), not all of which may be present in any given cell. Where we identify a primary minimum in the width profile, $\wmin$, we can also determine two local maxima, $\wstmax$ and $\wswmax$, corresponding respectively to the stalked and swarmer portions (Supplementary Fig.\ 2a). These values are approximately constant throughout the cell cycle, with $\mean{\wstmax(\phase)} > \mean{\wswmax(\phase)}$. However, at later times ($\phase>0.6$) the value of $\wswmax(\phase)$ increases until $\mean{\wstmax(\phase)} \approx \mean{\wswmax(\phase)}$. In some cases, additional secondary local minima are observed, $w^{st}_{min}$ and $w^{sw}_{min}$, corresponding respectively to the stalked and swarmer portions (Supplementary Fig.\ 2b). Although we note the value of these quantities for early times here (where they are approximately equal to their respective local maxima), these minima can only be determined with certainty at later times ($\phase>0.6$). There, we observe the presence of a secondary minimum in the stalked portions of most cells.

\begin{table}[!htp]
   \begin{tabular}{p{5.5em} p{6.5em} p{3em} l}
   	\hline\\[-1.5em]
      					& mean $\pm$ S.D.	& 			& $\phase $		\\
	\hline\\[-1.5em]
      $ \wmean $	& 0.74 $\pm$ 0.02	& $\mu$m	& $\left[0, 0.5\right]$	\\
      $ \wstmax $	& 0.85 $\pm$ 0.02	& $\mu$m	& $\left[0, 0.5\right]$	\\
      $ \wstmin $	& 0.84 $\pm$ 0.03	& $\mu$m	& $\left[0, 0.5\right]$	\\
      $ \wswmax $	& 0.83 $\pm$ 0.02	& $\mu$m	& $\left[0, 0.5\right]$	\\
      $ \wswmin $	& 0.82 $\pm$ 0.02	& $\mu$m	& $\left[0, 0.5\right]$	\\[0.5em]
      $ R $			& 4.44 $\pm$ 2.12	& $\mu$m	& $\left[0, 0.5\right]$	\\
      $ \Rst $			& 4.34 $\pm$ 2.50	& $\mu$m	& $\left[0, 0.5\right]$	\\
      $ \Rsw $		& 3.94 $\pm$ 2.42	& $\mu$m	& $\left[0, 0.5\right]$	\\[0.5em]
      $ \lst / \l $		& 54.3 $\pm$ 5.2		& \% 		& $\left[0, 1\right]$	\\
      $ \lsw / \l $		& 45.7 $\pm$ 5.2		& \% 		& $\left[0, 1\right]$	\\[0.5em]
      \hline\\[-1em]
   \end{tabular}
   \caption{\textbf{Average values of the cell shape parameters that are assumed constant over the specified time interval of $\phase$}. The time-average value of the width $\mean{\wmean}$ is smaller than both $\mean{\wstmin}$ and $\mean{\wswmin}$ because it accounts for the width at the primary minimum $\wmin$. Note the close correspondence between either width maximum and its corresponding minimum for $\phase<0.6$, as well as $\wstmax$ and $\lstmin$. Both width and radius of curvature for the swarmer portion are smaller than their respective values for the stalked portion.}
   \label{table:parameters}
\end{table}

We split each cell into stalked and swarmer portions according to the location of the primary minimum in the width profile. The radius of curvature ($R$) was calculated as the radius of the best-fit circle to the cell medial axis. We tested two methods of determining the radius of curvature: (I) fitting the whole cell to a circle or (II) fitting the stalked and swarmer portions separately. At early times in the cell cycle, (II) gives poorer results because of fewer data points. At later times in the cell cycle, (I) gives poorer results because the stalked and swarmer portions can indeed have differing radii of curvature, which we attribute to the alignment of the swarmer portion of the cell with the direction of fluid flow after the division plane has narrowed enough that it becomes mechanically decoupled from the stalked portion, i.e., like a flexible hinge. However, in the mean the results of (II) are equal to the value calculated by (I) for earlier times. Therefore, we use only data from method (I) but exclude later time points (Supplementary Fig.\ 2c). Because of the variation in the values of the calculated radius of curvature was large, such that we cannot define a reasonable arithmetic mean without arbitrarily filtering the dataset, we first averaged the corresponding unsigned curvature (equivalent to $\mean{R^{-1}}$) and then converted to radius of curvature (i.e., what we report as $\mean{R}$ is actually the harmonic mean, calculated as $\mean{R^{-1}}^{-1}$).

\begin{figure*}[!htp]
\includegraphics[scale=0.58]{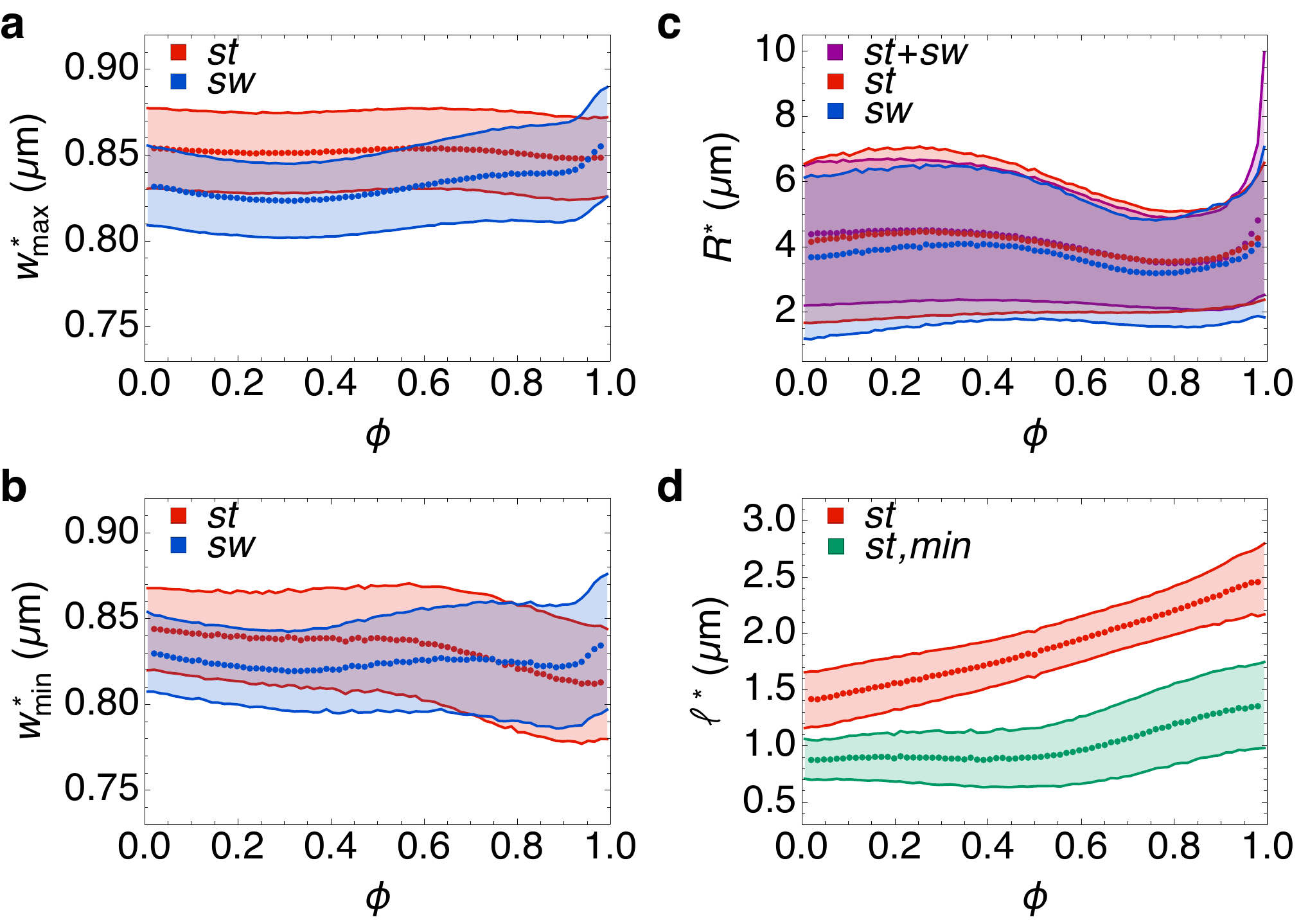}
\caption{\textbf{Unnormalized values of the cell shape parameters}. Time-dependence of \textbf{(a)} width maxima, and \textbf{(b)} width minima, of both stalked and swarmer portions. These data correspond to Fig. 5a in the main text. \textbf{(c)} Radius of curvature calculated for either the entire cell (black), the stalked portion only (red), or the swarmer portion only (blue). \textbf{(d)} Unnormalized length from stalked pole to either the primary (red) or secondary (green) width minima. These data correspond to Fig. 5b in the main text. Superscript $*$ refers to either $st$ or $sw$, respectively, as indicated in each figure legend.}
\label{fig:parameters}
\end{figure*}

The length was also split according to the locations of the local minima, into $\lst$ (distance along cell medial axis from stalked pole to $\wmin$), $\lsw$ (distance along cell medial axis from swarmer pole to $\wmin$), $\lstmin$ (distance along cell medial axis from stalked pole to $\wstmin$), and $\lswmin$ (distance along cell medial axis from swarmer pole to $\wswmin$). The value of $\lst$ and $\lstmin$ are compared in Supplementary Fig.\ 2d. Note that the length of the stalked portion $\lst$ grows exponentially, with the same time constant as the length $\l$ (as does $\lsw$, although it is not shown here for clarity), which is a necessary condition for the addition of peptidoglycan material along the entire length of the cell when the location of $\wmin$ is set at early times. At later times ($\phase>0.6$), the length from stalked pole to secondary minimum $\lstmin$ also starts to increase.

We relate the asymmetry in the length of stalked and swarmer portions at early times to asymmetries in the stalked and swarmer poles. The model predicts a linear relationship between the ratio of lengths $\lst / \lsw$ and the ratio of the tensions at either pole $\gamma^{st}_{p} / \gamma^{sw}_{p}$. We cannot directly measure the latter quantity, but from Laplace's law it is equal to the inverse ratio of the mean curvatures at either pole, which we approximate as the inverse of half the maximum pole width, assuming that the poles are hemispheres with diameter $\wstmax$ and $\wswmax$, respectively. Supplementary Fig.\ 3 shows scatter plots comparing $\wstmax / \wswmax$ to $\lst / \lsw$ for three different time intervals. The red best-fit line is shown only for Supplementary Fig.\ 3a, for which $R^{2} = 0.15$ (linear fits to all other plots produced values of $R^{2} < 0.05$). This line runs from the point $(1.0, 1.01)$, corresponding to the dashed blue curve in Fig. 3c of the main text (which is the case of a symmetric cell) to $(1.2, 1.04)$, corresponding to the red solid curve in Fig. 3d of the main text (which is the case of the average asymmetric \textit{C. crescentus} cell). Note that at times $\phase > 0.6$, any correlation disappears.

In order to quantify the error in our width profiles, we imaged a single field of view of 24 \textit{C. crescentus} cells perfused in complex medium at 31$^{\circ}$C at a frame rate of 5 frames per second (300 times faster than the frame rate used to acquire all other data), and calculated the splined contours for each cell. We focused in particular on a single ``dead'' (non-growing and non-dividing) cell, and found the root-mean-square deviation (RMSD) of nearest points along the cell contour between subsequent frames to be 12 nm. In addition, we found the RMSD of equivalent points along the width profile between subsequent frames to be 28 nm, or a 3.2\% pinch depth at the average value of $\wstmax$ = 0.85 $\mu$m.

\section*{Cell Shape Model}

\begin{figure*}[!htp]
\includegraphics[scale=0.58]{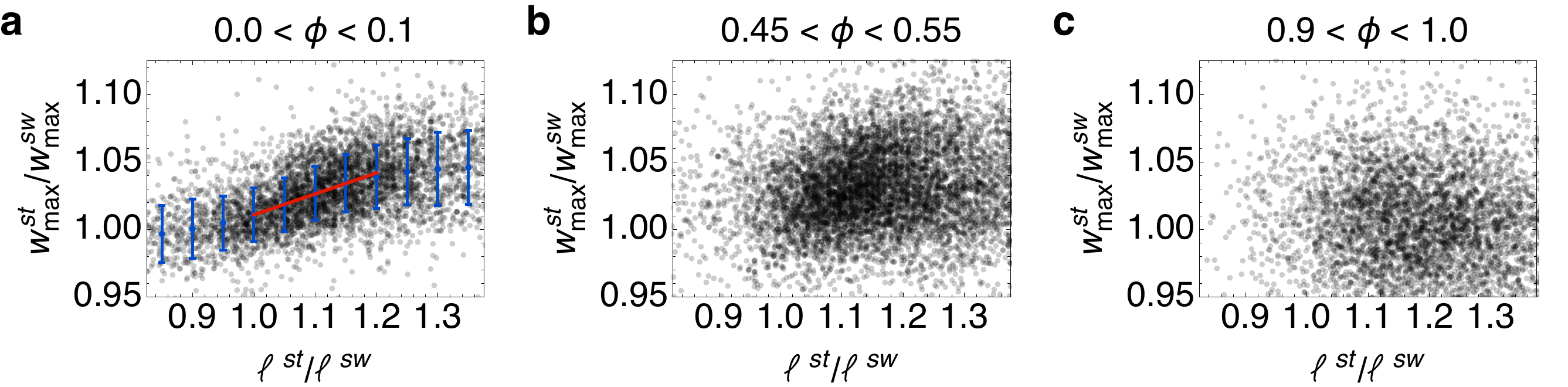}
\caption{\textbf{Ratios of maximum width at stalked:swarmer poles versus stalked:swarmer lengths}. The three columns delineate the time period from which data were taken: \textbf{(a)} first 10\% of the generation, \textbf{(b)} middle 10\% of the generation, and \textbf{(c)} last 10\% of the generation. Blue points show the average value of $\wstmax / \wswmax$ calculated by binning $\lst / \lsw$ into bins of width 0.05, with size of each error bar showing the standard deviation of the values within that bin. The best-fit line for $1 \leq \lst / \lsw \leq 1.2$ is indicated in red (coefficient of determination $R^{2} = 0.15$). This line runs from $(1.0, 1.01)$ to $(1.2, 1.04)$.}
\label{fig:endcaps}
\end{figure*}

\textbf{Cell Wall Mechanics}. The total energy $E$ for the bacterial cell wall is given as the sum of contributions from an active internal pressure $P$ driving cell volume ($V$) expansion, mechanical energy $E_\text{wall}$ in the cell wall, and the mechanical energy of interactions with cytoskeletal proteins $E_\text{proteins}$:
\begin{equation} 
E=-PV+ E_\text{wall} + E_\text{proteins}\;.
\end{equation}
The bacterial cell wall consists of a network of glycan strands cross-linked by peptide chains know as the peptidoglycan network. Growth occurs via the insertion of new peptidoglycan strands into the existing network along with the breaking of existing bonds due to turgor pressure induced stretching. We assume that elastic equilibrium is reached rapidly as compared to the rate of synthesis of new material~\Scite{goriely2008}. As a result of cell wall remodeling and irreversible elongation, growth can be understood as resulting from plastic deformations~\Scite{cosgrove1985,koch2001,boudaoud2003}. To understand the origin of the cell wall tension, $\gamma$, in the model introduced in the main text, we consider the cell wall as a thin elastic shell that deforms plastically when stretched beyond a maximum strain $\varepsilon_Y$, the yield strain. A thin shell has two modes of elastic deformations, bending and stretching~\Scite{landau1959}, such that $E_\text{wall}=E_\text{stretch} + E_\text{bend}$. In the limit of small thickness of the shell $h$ as compared to its radii of curvature, one can neglect the bending energy (that scales as $E_\text{bend}\sim h^3$) whereas the stretching energy $E_\text{stetch}$ is given by,
\begin{equation} 
E_\text{stretch}=\frac{h}{2}\int dA\ \sigma_{ij}\varepsilon_{ij}
\end{equation}
where $\sigma_{ij}$ is the mechanical stress tensor and $\varepsilon_{ij}$ is the strain tensor. As yield strain is reached at the onset of growth, we have $\varepsilon_{ij}=\varepsilon_Y\delta_{ij}$ (assuming isotropic stretching), where $\delta_{ij}$ is the Kronecker delta. Furthermore, assuming a Hookean constitutive relation for the stress tensor~\Scite{landau1959} 
\begin{equation}
\sigma_{ij}=\frac{Y}{1-\nu^2}\left[(1-\nu)\varepsilon_{ij}+\nu\varepsilon_{kk}\delta_{ij}\right]\;,
\end{equation}
where $Y$ is the Young's modulus and $\nu$ is the Poisson ratio, we have $E_\text{stretch}=\gamma A$, where,
\begin{equation}
\gamma=\frac{1}{2}\sigma_{ij}\varepsilon_{ij}=\frac{Yh\varepsilon_Y^2}{(1-\nu)}\;.
\end{equation} 
For a Gram-negative bacterial cell wall of thickness $h\sim 3$ nm, elastic modulus $Y\sim 40$ MPa~\Scite{boulbitch2000} and average yield strain $\varepsilon_Y\sim0.5$, the wall tension is estimated to be $\gamma=50$ nN$/\mu$m. While the actual value for the turgor pressure counteracting this tension can be contested, our choice for the numerical value for internal pressure $P$ can be justified using a simple mechanical argument. Radial force-balance dictates that in order to maintain an average cross-sectional radius $r \simeq 0.4$ $\mu$m, a cell wall with surface tension $\gamma \sim$ 50 nm/$\mu$m  has to balance an internal pressure of magnitude $P=2\gamma/r\simeq$ 0.25 MPa, which is numerically very close to our choice for $P=0.3$ MPa.

\textbf{Cytoskeletal bundle mechanics}. Next, we model the mechanical energy in the cell wall due to interactions with cytoskeletal proteins. Our minimalist approach considers two crucial protein bundles that are directly responsible for maintaining the shape of \textit{C. crescentus} cells. MreB protein bundles contribute an effective energy $E_\text{width}$, which favors a rod-like shape, and crescentin filament bundles contribute an energy $E_\text{cres}$, which favors a crescent-like shape. We thus have $E_\text{proteins}=E_\text{width}+E_\text{cres}$. MreB subunits form patchy filamentous bundles adherent to the cell wall and oriented perpendicular to the long axis of the cell~\Scite{garner2011,gahlmann2014}. The elastic energy stored in an adherent MreB subunit is given by
\begin{equation}
E_{\text{MreB},i}=\frac{k_\text{b}}{2}\ell_i \left(\frac{1}{r_i}-\frac{1}{R_m}\right)^2\;,
\end{equation}
where $i$ labels the subunit, $r_i$ is the circumferential radius of curvature of the cell wall where the subunit is attached, $\ell_i$ is the length of the subunit, $R_m$ is the intrinsic radius of curvature and $k_b$ is the bending rigidity of the associated MreB bundle. The total energy imparted by a collection of $N_m$ attached MreB subunits is given by $E_\text{width}=\sum_i^{N_m} E_{\text{MreB},i}$. Next, employing a continuum mean field assumption, we replace individual subunit lengths by their average length $\langle \ell \rangle \simeq 5$ nm~\Scite{colavin2014} and assume a uniform number density $\rho_m$ of MreB subunits in the cell surface to obtain
\begin{equation}
E_\text{width}=\frac{k_\text{m}}{2}\int dA \left(\frac{1}{r}-\frac{1}{R_m}\right)^2\;,
\end{equation}
where, $k_m=k_b\rho_m \langle \ell\rangle$ is the effective bending modulus due to MreB induced traction forces. The bending rigidity of an MreB bundle is given by $k_b=k_\text{MreB} n^\xi$, where $k_\text{MreB}$ is the flexural rigidity of MreB filaments (assumed to be similar to F-actin), $n$ is the number of MreB protofilaments per bundle, and $\xi$ is an exponent in the range 1--2 depending on the strength of crosslinking or bundling agents~\Scite{bathe2008}. MreB filaments in a bundle appear to have strong lateral interaction with negligible filament sliding, so we assume $\xi\simeq 2$. The diameter of an MreB protofilament is $\sim 4$ nm~\Scite{van2001} and the average width of an MreB bundle has been determined from super-resolution imaging to be in the range 60-90 nm~\Scite{reimold2013}, giving the estimate $n\sim$ 15-22. Estimating the surface number density of MreB subunits as $\rho_m\sim 5\times10^4$ $\mu$m$^{-2}$ and using $k_\text{mreb}\simeq 10^{-4}$ nN$\mu$m$^2$~\Scite{gittes1993}, we obtain $k_m$ in the range 5.6--12.1 nN$\mu$m.

Crescentin proteins form a cohesive bundled structure anchored to the sidewall of \textit{C. crescentus} cells. The energetic contribution due to crescentin is given by
\begin{equation}
E_\text{cres}=\frac{k_c}{2}\int ds\left(c(s)-\frac{1}{R_c}\right)^2\;,
\end{equation}
where $k_c$ is the bending rigidity, $s$ is an arc-length parameter, $c(s)$ is the longitudinal curvature of the cell wall and $R_c$ is the preferred radius of curvature of crescentin bundles. The bending rigidity of crescentin can be expressed as $k_c = Y_c I$, where $Y_c$ is the Young's modulus and $I$ is the area moment of inertia of the bundle (with width $w_c$) given by $I=\pi w_c^4/64$. Since crescentin is an intermediate filament homologue, we assume that $Y_c$ is similar to the Young's modulus of intermediate filament bundles given by $Y_c \sim 300$ MPa~\Scite{guzman2006}. Assuming $w_c \sim 0.1$ $\mu$m, we estimate the bending rigidity of a crescentin bundle to be $k_c \sim 1.5$ nN$\mu$m$^2$.

\textbf{Mean field model for cell shape and size dynamics}. As described in the main text, the total mechanical energy in the cell wall of \textit{C. crescentus} cells can be given as a sum of contributions from internal pressure, wall surface tension, mechanical energy of interactions with bundles of cytoskeletal proteins such as crescentin and MreB, and the constriction energy $E_\text{div}$ during cell division. In the mean field description, we neglect the contributions from $E_\text{div}$ and disregard any spatial variations in cell geometry. The mean field description is a good approximation of the cell shape dynamics for $\phase<0.5$, where the average width and the midline radius of curvature remains constant (Fig. 2 in the main text). We approximate the shape of \textit{C. crescentus} cells as the segment of a torus with radius of curvature $R(t)$, cross-sectional width $\wmean(t)$ and spanning angle $\theta(t)$. We also neglect the pole caps that are mechanically rigid and do not remodel during wall growth. The total energy is then given by
\begin{equation}
E[\wmean,R,\theta]=-P (\pi \wmean^2 R\theta)/4 + \gamma (\pi R \wmean \theta) + E_\text{width} + E_\text{cres}\;,
\end{equation}
where
\begin{equation}
E_\text{width}=\frac{k_m}{2}(\pi R w \theta)\left[(\wmean/2)^{-1}-R_m^{-1}\right]^2 \;,
\end{equation}
\begin{equation}
E_\text{cres}=\frac{k_c}{2}(R-\wmean/2)\theta\left[(R-\wmean/2)^{-1}-R_c^{-1}\right]^2\;.
\end{equation}
From the above expressions, we see that the total energy has the scaling form $E[\wmean,R,\theta]=\theta U[\wmean,R]$. According to Eq.~(1) in the main text, the dynamics of the radius of curvature $R$, the spanning angle $\theta$ and the width $w$ are given by
\begin{subequations}
\begin{gather}
\frac{1}{R}\frac{dR}{dt}=-\Phi_R \frac{\partial E}{\partial R} \;,\\
\frac{1}{\wmean}\frac{d\wmean}{dt}=-\Phi_w \frac{\partial E}{\partial \wmean}\;,\label{eq:wmean}\\
\frac{1}{\theta}\frac{d\theta}{dt}=-\Phi_\theta \frac{\partial E}{\partial \theta}\;, \label{eq:theta}
\end{gather}
\end{subequations}
where $\Phi_R$, $\Phi_\theta$ and $\Phi_w$ are the rate constants. The steady-state values for the radius of curvature of the centerline and the cross-sectional width is given by the solutions to $\partial_R U=0$ and $\partial_w U=0$. As shown in Supplementary Fig.\ 4a the energy density $U$ admits a stable absolute minimum in width, controlled by the stiffness parameter $k_m$ such that for lower values of $k_m$ (red curve in Supplementary Fig.\ 4a), $U$ does not have a minimum and the width grows in time rendering the rod-like shape unstable. Similarly, the parameter $k_c$ controls cell curvature, such that the energy density has a stable absolute minimum in $R$ beyond a critical stiffness $k_c$ (Supplementary Fig.\ 4b). At lower values of $k_c$ the cell does not have a stable radius of curvature and the energy is minimized by increasing $R$. This leads to straightening of the cell's medial axis. In the parameter range where the cell maintains a stable value for $\wmean$ and $R$, the energy density $U$ becomes a constant during cell elongation. From the growth law introduced in the main text, the spanning angle $\theta$ evolves in time according to the equation
\begin{equation}
\frac{d\theta}{dt}=-(\Phi_\theta  U) \theta(t)\;,
\end{equation}
such that the condition for exponential growth becomes $U<0$.

\begin{figure*}[!htp]
\includegraphics[scale=0.58]{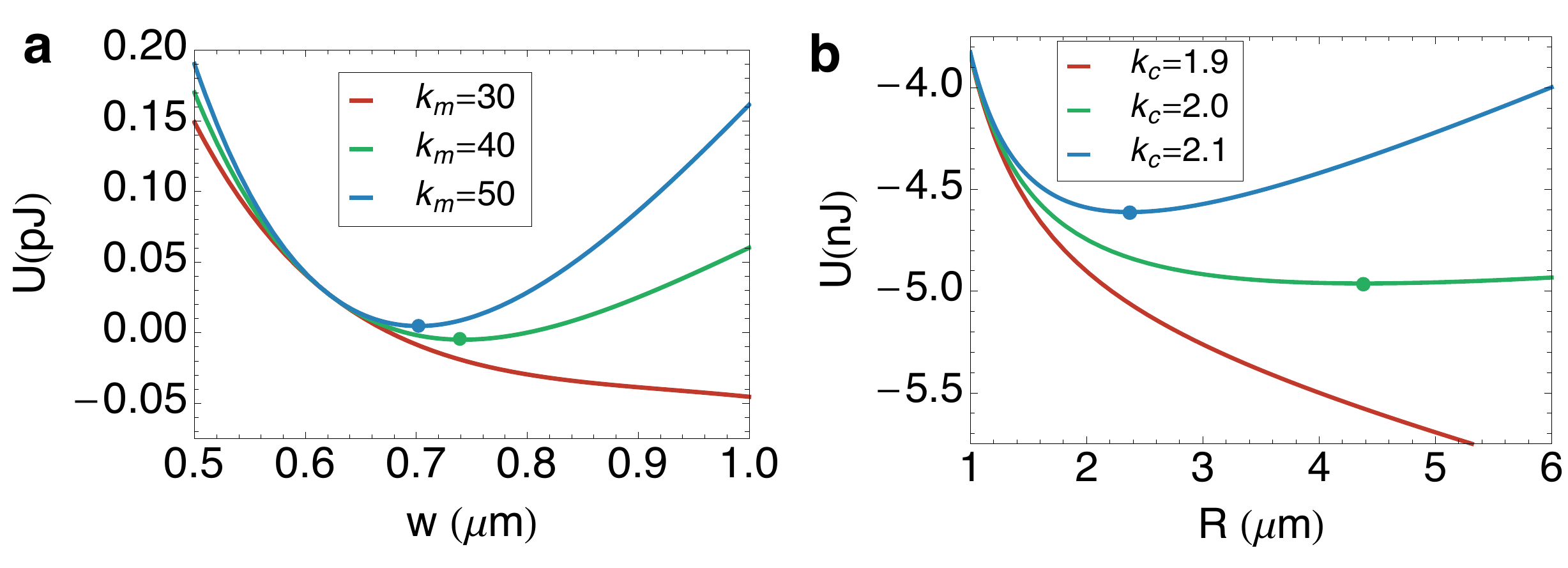}
\caption{\textbf{Dependence of the mechanical energy on cell width and radius of curvature}. \textbf{(a)} Energy density $U$ as a function of mean cell width $\wmean$ at various values of bending stiffness (in units of $nN\mu$m): $k_m=30$ (red), $k_m=40$ (green) and $k_m=50$ (blue), for a fixed radius of curvature $R=4.4\ \mu$m and $k_c=2$ nN$\mu$m$^2$. The absolute minima (if it exists) are indicated by solid circles. \textbf{(b)} $U$ as a function of radius of curvature $R$ at various values of crescentin stiffness (in units of $nN\mu$m$^2$): $k_c=1.9$ (red), $k_c=2.0$ (green) and $k_c=2.1$ (blue), for a fixed width $\wmean=0.74\ \mu$m and $k_m=40$ nN$\mu$m. Other parameters: $P=0.3$ MPa, $\gamma=50$ nN/$\mu$m, $R_c=0.5$ $\mu$m, $R_m=0.31$ $\mu$m.}
\label{fig:energy}
\end{figure*}

\textbf{Constriction dynamics}. As described in the main text, we assume that the shape of the constriction zone is given by two intersecting hemishperical segments with diameter $w_\text{max}$. The total surface area of the septum is given by $S(t)=\pi w_\text{max} \l_s(t)$, where $\l_s(t)$ is the total length of the spherical segments (see Supplementary Fig.\ 5a). Using elementary geometry, we obtain the following relation between $\l_s(t)$ and the minimum width, $w_\text{min}(t)$:
\begin{equation}\label{eq:wmin}
w_\text{min}(t)=w_\text{max}\sqrt{1-(\l_s(t)/w_\text{max})^2}\;.
\end{equation}
If septal growth occurs by addition of new peptidoglycan strands while maintaining the curvature of the spherical segments, then $\l_s(t)$ is the shape variable controlling the growth of septal surface area $S$ as well as the constriction dynamics of $w_\text{min}(t)$. To describe the dynamics of $\l_s(t)$, we assume that initial phase of elongation occurs with a velocity $v_0$, and thereafter growth follows an exponential law with rate $\kappa_d$,
\begin{equation}
\frac{d\l_s}{dt}=v_0 + \kappa_d \l_s(t) \;.
\end{equation}
Multiplying both sides of the equation by $\pi w_\text{max}$, we derive the growth dynamics of the surface area S(t),
\begin{equation}
\frac{dS}{dt}=\kappa_0 + \kappa_d S(t) \;,
\end{equation}
where $\kappa_0=\pi w_\text{max}v_0$. Having determined the dynamics of $w_\text{min}$, one can evaluate the time-dependence of the average width $\wmean$ defined as, $\wmean (t)=\frac{1}{\l(t)}\int_0^{\l(t)} w(u,t) du$, where $u$ is the coordinate along the centerline. Using the simplified geometry of the constricting cell, given by two toroidal segments (peripheral regions) connected by two intersecting hemispheres (septal region), one gets,
\begin{equation}\label{eq:wav}
\wmean(t)=\frac{1}{l(t)}\left[ \wmean(\l(t)-\l_s(t)) + 2\int_0^{\l_s(t)/2} w_s(u,t) du\right]\;,
 \end{equation}
where $w_s(u)=w_\text{max}\sqrt{1-(2u/w_\text{max})^2}$, is the width in the septal region and $\wmean$ is the average width of the stalked and swarmer components as determined by Eq.~\eqref{eq:wmean}. As shown in Fig. 2c in the main text (blue solid line), Eq.~\eqref{eq:wav} is in excellent agreement with the experimental data and captures the dip in $\wmean(t)$ seen for $\phase>0.5$. Fig. 2b in the main text also shows that the radius of curvature of the centerline ($R(t)$) drops for $\phase>0.5$ when cell constriction is prominent. Therefore we examine the role of constriction dynamics on the time-dependence of the centerline radius of curvature. 

The centerline spans a distance $\l_s(t)$ and an angle $\theta_s(t)$ in the septal region, as shown in Supplementary Fig.\ 5a. Consider a small segment $\delta \l_s(t)$ along the centerline with a local radius of curvature $R$ and spanning angle $\delta\theta_s$. We then have the relation $\delta \l_s(t) = R(t)\delta \theta_s(t)$. This leads to the following identity,
\begin{equation} \label{eq:kinetic1}
\frac{1}{\delta l_s}\frac{ d(\delta l_s)}{dt}=\frac{1}{R}\frac{dR}{dt} + \frac{1}{\delta\theta_s}\frac{d(\delta \theta_s)}{dt}\;.
\end{equation}
Now the geometry of the septal region (Supplementary Fig.\ 5a) directly relates the rate of increase in the spanning angle $\delta\theta_s(t)$ to the rate of drop in the tangent angle ($\psi(t)$) at the constriction site, $\frac{1}{\delta\theta_s}\frac{d(\delta \theta_s)}{dt}=-\frac{1}{\delta\psi}\frac{d(\delta \psi)}{dt}$. Furthermore, using the relations, $w_\text{max}\sin{\psi(t)}=w_\text{min}(t)$ and $w_\text{max}\cos{\psi(t)}=l_s(t)$, we get $\delta\psi(t)=\delta w_\text{min}(t)/l_s(t)=-\delta l_s(t)/w_\text{min}(t)$. The last equality follows after variations of Eq.~\eqref{eq:wmin}. As a result we have the following kinetic relation,
\begin{equation}\label{eq:kinetic2}
\frac{d(\delta \psi)/dt}{\delta \psi}=\frac{d w_\text{min}/dt}{w_\text{min}}-\frac{d(\delta l_s)}{\delta l_s}=-\frac{d(\delta \theta_s)/dt}{\delta \theta_s}\;.
\end{equation}
Combining the identities in Eq.~\eqref{eq:kinetic1} and \eqref{eq:kinetic2} we get, $R^{-1}dR/dt=w_\text{min}^{-1}d w_\text{min}/dt$, showing that the radius of curvature drops at the same rate as $w_\text{min}(t)$ shrinks.

\begin{figure*}[!htp]
\includegraphics[scale=0.58]{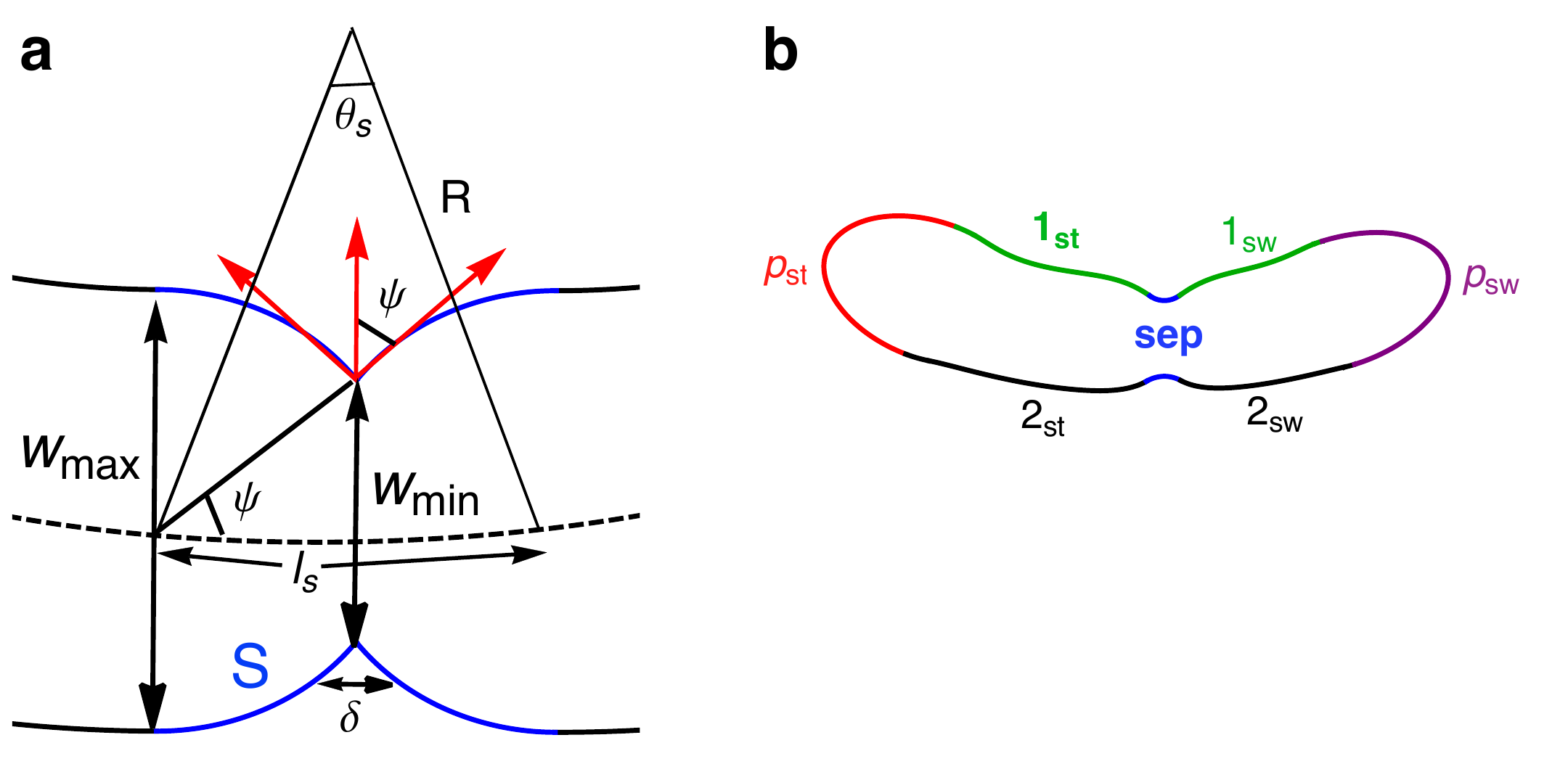}
\caption{\textbf{Geometry of a constricting cell contour}. \textbf{(a)} Schematic of the constriction zone of a dividing cell where the dashed line indicates the centerline with length $l_s$, radius of curvature $R$ and spanning angle $\theta_s$. S (blue) denotes the septal cell wall, $\psi$ is the tangent angle at the constriction site and $\delta$ is the width of the division ring. \textbf{(b)} Compartmentalizing the cell contour into the (1) stalked pole $p_{st}$ (red), (2) swarmer pole $p_{sw}$ (purple) (3) upper wall $1_{st/sw}$ (with crescentin bundle attached, green), (4) Lower wall $2_{st/sw}$ (black) and the (5) Septal region (blue).}
\label{fig:constriction}
\end{figure*}

\textbf{Contour Model for cell shape}. To quantitatively capture the experimentally observed spatial variations in cell shape we study an effective two-dimensional contour model for the cell shape. This approach facilitates closer comparison with the two-dimensional splined contours obtained from our experimental data. The contour description of the rod-like cell shape negelects circumferential variations in cell geometry. The model incorporates non-uniform materials properties and mechanical constraints across the cell wall, with the poles and the septal region being mechanically stiffer than the rest of the cell wall. By exploiting the rod-like geometry, one can use the centerline curve to divide the contour into two parts, the \textit{upper} and the \textit{lower} cell wall. As shown in Supplementary Fig.\ 5b we further subdivide the cell contour into the stalked and swarmer poles ($p_{st/sw}$), the upper ($1_{st/sw}$) and lower cell wall ($2_{st/sw}$), and the septal region (sep).

We parametrize the instantaneous shape of the cell contour using the two-dimensional centerline curve ${\bf R}(u,t)$, the length of the centerline $l(t)$ and the width $w(u,t)$, where $u$ is the absolute distance along the centerline from the stalked pole such that $u\in [0,l(t)]$. If $\hat{{\bf n}}$ denotes the outward unit normal vector on the centerline, the curves defining the upper and lower parts of the cell contour, ${\bf r}_\pm(u,t)$, are given by the relation, ${\bf r}_\pm={\bf R}\pm w_\pm\hat{{\bf n}}$, where $w_\pm(u)$ represent the perpendicular distances of the top and bottom curves from the centerline. The total cell width is then given by $w(u,t)=w_+(u,t) + w_-(u,t)$.

It is convenient to switch to polar coordinates, where the shape of the cell contour is given by the re-parametrized curve ${\bf r}_\pm(\varphi,t)$, where $0<\varphi<\theta(t)$ is the angular coordinate spanning the centerline, which can be approximated as the arc of a circle with radius $R(t)$. Since the ratio $w_\pm/R\ (\simeq 0.1)$ is small at all times, one can approximate the local curvature as
\begin{equation}\label{eq:curvature}
c_\pm(\varphi,t)=\frac{1}{R(t)}\left(1-\frac{w_\pm'(\varphi,t)+w_\pm''(\varphi,t)}{R(t)}\right) +\mathcal{O}(w_\pm^2, w_\pm'^2, w_\pm''^2)\;,
\end{equation} 
where prime denotes derivative with respect to $\varphi$ and the subscripts $\pm$ represent the upper and the lower part of the cell contour respectively. Furthermore, in the linear regime, the differential arc length can be approximated as $du\simeq Rd\varphi$. The dynamics of the shape parameters $l(t)$, $\theta(t)$, and $R(t)$ are determined from the kinetic law in Eq. (1) of the main text.

The instantaneous width profile $w(u,t)$ results from minimizing the total energy functional, which leads to the following shape equation,
\begin{equation}\label{eq:shape}
P-\gamma(u) c_\pm(u,t) + f_\text{width}(u,t) + f_\text{cres}(u,t)=0\;,
\end{equation}
where, $\gamma(u)$ is the tension on the cell contour and $f_\text{width}(u,t)$ and $f_\text{cres}(u,t)$ are the linear force densities on the contour due to maintenance of width and the crescent shape, respectively.
In the contour model we simplify the energetic contribution due to maintenance of width (acting on the sections $1_{st/sw}$ and $2_{st/sw}$) as
\begin{equation}\label{eq:mreb}
E_\text{width}=\frac{K_m}{2}\int du \left(w_\pm-R_m\right)^2\;,
\end{equation}
where the elastic constant $K_m$ depends on the bending rigidity $k_m$ introduced in the main text as $K_m=k_m/R_m^4$. This linear approximation holds if $(w_\pm-R_m)/R_m\ll 1$. From our data and mean field model fits we get $(w_\pm-R_m)/R_m\simeq 0.15$. The resultant force density is given by $f_\text{width}(u,t)=-K_m\left(w_\pm(u,t)-R_m\right)$.

The bending energy induced by crescentin protein bundles anchored onto the cell wall (regions $1_{st/sw}$) is given by
\begin{equation}\label{eq:cres}
E_\text{cres}=\frac{k_c}{2}\int ds \left[c_+(u)-c_0\right]^2\;,
\end{equation}
where $c_0=1/R_c$ is the spontaneous curvature of the crescentin bundle, and s is the arc-length parameter along the upper part of the cell contour ($1_{st/sw}$) which is related to $u$ as $ds=du-w_+d\varphi$. To obtain the force density we consider an infinitesimal deformation, $\delta r$, of the upper contour as ${\bf r}_+\rightarrow {\bf r}_+ +  {\bf n}_+ \delta r$, where ${\bf n}_+$ is the outward unit normal. Accordingly the curvature and the differential arc-length changes, $c_+\rightarrow c_+ + \delta c_+$ and $ds \rightarrow ds + \delta (ds)$, where $\delta c_+=c_+^2 \delta r  + d^2(\delta r)/ds^2$ and $\delta (ds) =-c_+ ds$~\Scite{mumford1994}. The resultant force density is obtained after variations of the energy functional, $\delta E_\text{cres}=\int f_\text{cres} \delta r ds$, where $f_\text{cres}\delta r ds=\delta[(c_+-c_0)^2 ds]$. This leads to the following non-linear force contribution:
\begin{equation}
f_\text{cres}=\frac{k_c}{2}\left(2\frac{d^2c_+}{ds^2}+c_+^3-c_+c_0^2\right)\;.
\end{equation} 
Using Eq.~\eqref{eq:curvature}, $f_\text{cres}$ can be linearized and expressed using the angular coordinate $\varphi$ as
\begin{equation}
f_\text{cres}(u,t)=\frac{k_c}{R^4}\left[ - w_+''''(u,t) + \frac{1}{2}(c_0^2R^2-5) w_+''(u,t) + \frac{1}{2}(c_0^2R^2-3) w_+(u,t) + R(1-c_0^2R^2)\right]\;.
\end{equation}

\textbf{Compartmentalizing the cell contour}. (1) Pole caps: The cell poles are assumed to be mechanically inert in the sense that they do not interact with the active cytoskeletal proteins such that $f_\text{width}=f_\text{cres}=0$. The mechanical forces acting on the cell poles come from turgor pressure $P$, and the tension $\gamma_p$. The shape of the cell poles are then described by the two-dimensional Laplace's law,
\begin{equation}
\gamma_p^{st,sw}c_\pm(u,t)=P\;,
\end{equation}
where the superscripts $st$ and $sw$ denote respectively the stalked and swarmer poles. 

(2) Upper cell contour ($1_{st/sw}$): The region $1_{st/sw}$ in the upper cell contour obeys the force-balance equation:
\begin{equation}
P-\gamma c_+(u,t) + f_\text{width}(u,t) + f_\text{cres}(u,t)=0\;.
\end{equation}
At the endpoints of the segments $1_{st/sw}$ we impose the boundary condition that the curvature $c$ must equal the longitudinal curvature of the poles. 

(3) Bottom cell contour ($2_{st/sw}$): In the absence of crescentin, the region $2_{st/sw}$ in the bottom cell contour obeys the force-balance equation:
\begin{equation}
P-\gamma c_-(u,t) + f_\text{width}(u,t)=0\;.
\end{equation}

(4) Septal region: We incorporate the effect of constriction in the cell shape equation by imposing the constraint (boundary condition) that $w(\lst(t),t)=w_\text{min}(t)$, where $w_\text{min}(t)$ evolves according to the kinetics described in Eq. (6) and (7) of the main text. Furthermore, the curvatures at the end points of the septal segments must conform to the curvature of the newly formed poles. Subject to these boundary conditions, the width profile in the upper septal region ($\lst-\l_s/2<u<\lst+\l_s/2$) is given by,
\begin{equation}
P-\gamma_s c_+(u,t) + f_\text{width}(u,t) + f_\text{cres}(u,t)=0\;,
\end{equation}
with a surface tension ($\gamma_s$ ) of the newly formed poles chosen to be much higher than the peripheral region, $\gamma_s \simeq 4.5 \gamma$. The contour below the centerline obeys the equation,
\begin{equation}
P-\gamma_s c_-(u,t) + f_\text{width}(u,t)=0\;.
\end{equation}

The governing cell shape equation given in Eq.~\eqref{eq:shape}, is then solved numerically in each part of the cell contour with matching boundary conditions in $w_\pm(u,t)$ and its derivatives. 

\section*{Supplementary References}

\end{document}